\documentclass[fleqn,usenatbib]{mnras}

\usepackage{newtxtext,newtxmath}
\usepackage{multirow}

\usepackage[T1]{fontenc}
\usepackage{ae,aecompl}

\usepackage{graphicx}	
\usepackage{amsmath}	

\newcommand{\tabhead}[1]{\textbf{#1}}

\newcommand{\comment}[1]{}

\usepackage[flushleft]{threeparttable}

\usepackage[dvipsnames]{xcolor}
\usepackage{soul}

\usepackage{tablefootnote}

\newcommand{\targ}{GK\,Per}
\newcommand{\kms}{ km s$^{-1}$}

\title[Stellar masses in GK Per]{The intermediate polar cataclysmic variable GK Persei 120 years after the nova explosion: a first dynamical mass study}

\author[A. \'Alvarez-Hern\'andez et al.]
       {A. \'Alvarez-Hern\'andez$^{1,2}$\thanks{E-mail: ayozeav@iac.es}, 
        M. A. P. Torres$^{1,2}$, P. Rodr\'\i guez-Gil$^{1,2}$, T. Shahbaz$^{1,2}$, \newauthor G. C. Anupama$^{3}$, K. D. Gazeas$^{4}$, M. Pavana$^{3,5}$, A. Raj$^{6}$, P. Hakala$^{7}$, G. Stone$^{8}$, \newauthor S. Gomez$^{9}$, P. G. Jonker$^{10,11}$, J.-J. Ren$^{12}$, G. Cannizzaro$^{10,11}$, I. Pastor-Marazuela$^{13,14}$, \newauthor W. Goff$^{15}$, J. M. Corral-Santana$^{16}$,  R. Sabo$^{17}$\\
\\
	$^1$Instituto de Astrof\'\i sica de Canarias, E-38205 La Laguna, Tenerife, Spain\\
        $^2$Departamento de Astrof\'\i sica, Universidad de La Laguna, E-38206 La Laguna, Tenerife, Spain\\
        $^3$Indian Institute of Astrophysics, 560 034 Bangalore, India\\
        $^4$Section of Astrophysics, Astronomy and Mechanics, Department of Physics, National and Kapodistrian University of Athens, GR-15784 Zografos, Athens, Greece\\
        $^5$Department of Physics, Pondicherry University, Puducherry 605014, India\\
        $^6$Dept. of Physics \& Astrophysics, University Road, University Enclave, Delhi 110007, India\\
        $^{7}$Finnish Centre for Astronomy with ESO (FINCA), Quantum, University of Turku, FI-20014, Turku, Finland\\
        $^8$First Light Observatory Systems, 9 Wildflower Way, Santa Fe, NM 87506 USA\\
        $^9$Center for Astrophysics \textbar{} Harvard \& Smithsonian, 60 Garden Street, Cambridge, MA 02138, USA\\
        $^{10}$SRON, Netherlands Institute for Space Research, Sorbonnelaan 2, 3584 CA, Utrecht, The Netherlands\\
        $^{11}$Department of Astrophysics/ IMAPP, Radboud University, Heyendaalseweg 135,6525 AJ, Nijmegen, The Netherlands\\
        $^{12}$CAS Key Laboratory of Space Astronomy and Technology, National Astronomical Observatories, Chinese Academy of Sciences, Beijing 100101,\\ People's Republic of China\\
        $^{13}$Anton Pannekoek Institute, University of Amsterdam, Postbus 94249, 1090 GE Amsterdam, The Netherlands\\
        $^{14}$ASTRON, the Netherlands Institute for Radio Astronomy, Oude Hoogeveensedijk 4, 7991 PDDwingeloo, The Netherlands\\
        $^{15}$American Association of Variable Star Observers, 49 Bay State Rd. Cambridge, MA 02138\\
        $^{16}$European Southern Observatory, Alonso de C\'ordova 3107, Vitacura, Casilla 19001, Santiago de Chile, Chile\\
        $^{17}$American Association of Variable Star Observers, 1344 Post Dr. Bozeman, MT 59715}

\date{Accepted XXX. Received YYY; in original form ZZZ}

\pubyear{2020}

\begin{document}
\label{firstpage}
\pagerange{\pageref{firstpage}--\pageref{lastpage}}
\maketitle

\begin{abstract}
We present a dynamical study of the intermediate polar and dwarf nova cataclysmic variable \targ\ (Nova Persei 1901) based on a multi-site  optical spectroscopy and $R$-band photometry campaign. The radial velocity curve of the evolved donor star has a semi-amplitude $K_2=126.4 \pm 0.9 \, \mathrm{km}\,\mathrm{s}^{-1}$ and an orbital period $P=1.996872 \pm 0.000009 \, \mathrm{d}$. We refine the projected rotational velocity of the donor star to $v_\mathrm{rot} \sin  i  = 52 \pm 2 \, \mathrm{km}\,\mathrm{s}^{-1}$ which, together with $K_2$, provides a donor star to white dwarf mass ratio $q=M_2/M_1=0.38 \pm 0.03$. We also determine the orbital inclination of the system by modelling the phase-folded ellipsoidal light curve and obtain $i=67^{\circ} \pm 5^{\circ}$. The resulting dynamical masses are $M_{1}=1.03^{+0.16}_{-0.11} \, \mathrm{M}_{\odot}$ and $M_2 = 0.39^{+0.07}_{-0.06} \, \mathrm{M}_{\odot}$ at $68$ per cent confidence level. The white dwarf dynamical mass is compared with estimates obtained by modelling the decline light curve of the $1901$ nova event and X-ray spectroscopy. The best matching mass estimates come from the nova light curve models and an X-ray data analysis that uses the ratio between the Alfv\'en radius in quiescence and during dwarf nova outburst.
\end{abstract}

\begin{keywords}
 accretion, accretion discs -- binaries: close -- novae, cataclysmic variables -- stars: individual: GK Per (Nova Persei 1901)
 \end{keywords}



\section{Introduction} \label{intro}

Cataclysmic variables (CVs) are binary systems where a non-degenerate 
star fills its Roche lobe and transfers matter towards an accreting white dwarf \citep[WD;][see \citealt{warner-libro} and references therein]{kraft-64}. For a weakly magnetic WD, the mass from the donor star is accreted onto the surface via an accretion disc. In magnetic CVs, however, the magnetic field is strong enough to dominate at least part of the accretion flow. Polars are the most extreme magnetic CVs: the strong WD magnetic field ($B \gtrsim 10^7 \, \mathrm{G}$) prevents the formation of an accretion disc and forces the transferred material to follow the field lines onto one or both poles of the WD (\citealt{chanmugam-77}; see \citealt{cropper-90} for a review). In contrast, in intermediate polars (IPs) the magnetic field is only able to take control over the transferred plasma in close proximity to the WD \citep{patterson-1994}. In these systems, the accretion disc is truncated at a certain radius from the WD and the disc accretion flow is funneled from there to its magnetic poles along the field lines. A remarkable difference between both types of magnetic CV is the degree of synchronization of the WD spin with the orbit: in IPs the spin period is usually significantly shorter than the orbital period, while both periods are nearly equal for most polars \citep[see e.g.][]{norton-2004}.

GK Per was discovered as a nova on 1901 February 22 by Scottish amateur astronomer Thomas David Anderson \citep{williams-1901}. It peaked at a visual apparent magnitude of $0.2$~mag. After years of irregular fluctuations in brightness with amplitudes up to $1.5$ mag and several dozens of days duration, in $1948$ it reached a quiescence state ($m_V \simeq 13$~mag) and started to show $1-3$-mag dwarf nova outbursts that typically last $50$~d and recur about every three years \citep{hudec-1981,bianchini-1982,sabbadin-83,simon-2002}.

\cite{crampton-1986} reported a binary orbital period of nearly 2~d and \cite{watson-1985} found a WD spin period of $351\,\mathrm{s}$ in the modulation of the hard X-ray emission, thus confirming the IP nature of \targ. The spin period also modulates the $U$-band flux \citep{patterson-1991} and the equivalent width of the emission lines at optical wavelengths \citep{garlick-94,reinsch-94}. $J$-band circular polarimetry of \targ\ in quiescence is consistent with a null detection \citep{stockman-92}. However, the intensity of the WD magnetic field is estimated at $B \sim 10^5$\,G from X-ray spectral modelling \citep{wada-2018}.

The geometrical, kinematic and physical properties of the nova shell in \targ\ as well as its interaction with its surroundings have been studied in detail at different frequencies \citep[and references therein]{seaquist-89, scott-94, anupama-2005, liimets-2012}.  In particular, far-infrared observations showed that the nova shell is embedded in an ancient, possibly bipolar planetary nebula centred on the binary and extending $\approx 17$ arcmin to the NW and SE \citep{dougherty-96}. At the time of its discovery, this nebula was interpreted as being the remnant of the binary common envelope phase \citep{bode-87}. However, ejecta presumably from the WD progenitor star expelled during a second asymptotic giant branch phase (and thus a second common envelope event), triggered by a period of high mass transfer rate from the donor star ($>3 \times 10^{-7} \, \mathrm{M}_{\odot} \, \mathrm{yr}^{-1} = 1.9 \times 10^{19} \, \mathrm{g} \, \mathrm{s}^{-1}$), was proposed as a more likely origin for the nebula \citep{dougherty-96}.

Several spectroscopic classifications of the donor star in \targ\ have been reported by different authors: K2~V-IVp \citep{kraft-64,gallagher-74}, K0~III-IV \citep{crampton-1986}, K2--3~V \citep{reinsch-94} and K1~IV \citep[][hereinafter MR02]{moralesrueda}. MR02 presented a radial velocity study of the donor star building on similar work by \cite{kraft-64}, \cite{crampton-1986} and \cite{reinsch-94} that provided an orbital period $P=1.9968 \pm 0.0008$\,d, a radial velocity semi-amplitude $K_2=120.5 \pm 0.7$\,\kms\ and a systemic velocity $\gamma=40.8 \pm 0.7$\,\kms. MR02 also reported an estimate of the projected rotational velocity of the donor star on the line of sight of the observer ($v_\mathrm{rot} \sin  i = 61.5 \pm 11.8$\,\kms) and a donor-to-WD mass ratio $q = 0.55 \pm 0.21$. In obtaining these values they used optical spectra with $\approx 120$\,\kms\ full-width at half-maximum (FWHM) resolution. \cite{harrison-2015} obtained $v_\mathrm{rot} \sin  i = 55 \pm 10$\,\kms\ from near-infrared spectra with $\approx 12$~\kms\ FWHM spectral resolution. 

Precise dynamical masses of the two stars in \targ\ have never been determined because the orbital inclination has remained largely unconstrained. The absence of eclipses in the light curves at optical wavelengths suggested an inclination $i < 73^{\circ}$ \citep{reinsch-94}, which translates to lower limits on the masses of the WD and the donor star of $M_{1} > 0.87 \pm 0.24$\,M$_{\odot}$ and $M_{2} > 0.48 \pm 0.32$\,M$_{\odot}$, respectively (MR02). WD masses of $M_{1}=1.15 \pm 0.05 \, \mathrm{M}_{\odot}$ and $M_{1}=1.22 \pm 0.10 \, \mathrm{M}_{\odot}$ have been derived from modelling of the nova light curve by \cite{hachisu-2007} and \cite{shara-18}, respectively. In addition, estimates of the WD mass ranging from $0.52^{+0.34}_{-0.16}$ to $1.24 \pm 0.10$\,M$_{\odot}$ have been obtained through modelling of X-ray spectra (see Section~\ref{sec-discuss} for details).

In this work, we present the first dynamical study of \targ\ that yields reliable masses for the WD and the donor star. The paper is structured as follows: in Section~\ref{sec-obs} we describe the time-resolved optical photometry and spectroscopy observations and the data reduction. From the analysis of the absorption lines of the donor star we obtain its radial velocity curve (Section~\ref{sec-k2}), constrain its spectral type (Section~\ref{sec-spectralclassification}) and determine its rotational broadening (Section~\ref{sec-vsini}). In Section~\ref{sec-lightcurve} we present the modelling of the $R$-band light curve,  which provides the orbital inclination for the first time. The stellar dynamical masses are then obtained using the measured quantities. These are discussed in Section~\ref{sec-discuss}, where we also compare the WD dynamical mass with the available estimates from X-ray spectral fitting and modelling of the nova light curve. Finally, we draw our conclusions in Section~\ref{sec-conclusions}.

\section{Observations and data reduction}\label{sec-obs}
The 1.9968-d orbital period of \targ\ makes a given orbital phase occur 4.6 min earlier every next orbital cycle. In addition, during a typical 10-h observing night only 20 per cent of the orbit can be covered. Thus, the fact that the orbital period is close to an integer number of days precludes ground-based observers at a single location from achieving entire photometric coverage of the orbit in contemporaneous nights. This makes the light curve strongly susceptible to aperiodic night-to-night accretion variability. To overcome this difficulty and thus obtain the full ellipsoidal modulation produced by the donor star, we performed multi-site photometry between 2017 December and 2018 January. We also took multi-site spectroscopy during 2017--2019 to improve the light curve modelling and obtain a full dynamical determination of the system parameters.

In this Section, we describe all the collected data sets. Tables~\ref{tab:observaciones_espectroscopia} and~\ref{tab:observaciones_fotometria} summarize the spectroscopic and photometric observations, respectively. Note that we have adopted orbital phase $0$ as the moment of inferior conjunction of the donor star.

\subsection{Spectroscopy}\label{subsec:spectroscopy}
The optical spectroscopy data of \targ\ were obtained in 2017--2019 using four telescopes. We planned the observations in order to cover the orbital phases that better define the radial velocity curve ($0.25$ and $0.75$) and $v_\mathrm{rot} \sin  i$. The spectral resolution of all our data sets was slit limited. Only the seeing of the 2019 September 7 WHT data ($0.7-0.8$~arcsec) was slightly less than the slit width (0.8~arcsec), but this was checked to have a negligible influence on the results. Except where indicated, in each observing run we observed the spectral templates HD~20165 and HR~2556, which are classified as K1 V \citep{koen-2010} and K0 III--IV \citep{luck-2015} stars and have low intrinsic radial velocities of $-16.92 \pm  0.19$~$\mathrm{km}\,\mathrm{s}^{-1}$ and $27.68 \pm  0.17$~$\mathrm{km}\,\mathrm{s}^{-1}$, respectively \citep{gaia}. Their rotational broadenings are also small: $1.6$~$\mathrm{km}\,\mathrm{s}^{-1}$ for HD~20165 \citep{brewer-2016} and $4.3$~$\mathrm{km}\,\mathrm{s}^{-1}$ for HR~2556 \citep{luck-2015}.

\subsubsection{Himalayan Chandra Telescope}\label{subsec-hct}

The first data set was taken with the 2-m Himalayan Chandra Telescope (HCT) located in the Indian Astronomical Observatory, Saraswati Mount, India. The Hanle Faint Object Spectrograph Camera (HFOSC) was used with grism $\# 8$ and a 0.77-arcsec slit width. This instrumental configuration provided spectra in the wavelength range $5120-9310$~\AA\ with a dispersion of 1.27~\AA \,~$\mathrm{pix}^{-1}$ and a FWHM spectral resolution of 5.6 \AA \, (equivalent to $\simeq 270$ $\mathrm{km}\,\mathrm{s}^{-1}$ at $6300$~\AA). We took 20, 24 and 21 spectra using exposure times between 900 and 1200~s on the nights of 2017 December 6, 7 and 8, respectively. FeNe calibration arc lamps were taken often. According to the ephemeris (Section~\ref{sec-k2}), these observations covered time intervals near the quadratures of the orbit. The seeing measured from the spectral traces was $1.7-3.6$~arcsec during the first night, $1.7-3.2$~arcsec on the second night and $1.7-2.6$~arcsec on the last night.

\subsubsection{Nordic Optical Telescope}\label{subsec-not}

We took a second data set with the 2.56-m Nordic Optical Telescope (NOT) sited in the Observatorio del Roque de los Muchachos on La Palma, Spain.  We used the Alhambra Faint Object Spectrograph and Camera (ALFOSC) with grism $\# 8$ and a 0.5-arcsec slit width. This setup yields a wavelength coverage $5680-8580$~\AA, a dispersion of 1.41 \AA\ $\mathrm{pix}^{-1}$ and a FWHM spectral resolution of 3.5 \AA\ (equivalent to $\simeq 170$ $\mathrm{km}\,\mathrm{s}^{-1}$ at 6300 \AA). These observations were conducted on the nights of 2017 December 8, 9 and 10, when we obtained 30, 21 and 16 spectra, respectively. The exposure time varied between 600 and 900~s. Spectra of HeNe + ThAr calibration arc lamps were taken after each target exposure. This data set covered orbital phase ranges around 0 and 0.5. The seeing was $1.1-2.1$, $1.0-1.2$ and $0.9-1.5$~arcsec on the first, second and third night, respectively.

\subsubsection{William Herschel Telescope}

In order to obtain further radial velocities and to measure the rotational broadening of the absorption lines of the donor star, we used the Intermediate-dispersion Spectrograph and Imaging System (ISIS) attached to the 4.2-m William Herschel Telescope (WHT), also located in the Observatorio del Roque de los Muchachos. We took a total of 24 optical spectra during seven nights between 2018 December 1 and 2019 September 13 using the R600R and the R1200R gratings with different slit widths and central wavelengths (see Table~\ref{tab:observaciones_espectroscopia}). The FWHM spectral resolutions at $6300$~\AA\ were $\leq 61$ and $\leq 36$ \kms\ for the R600R and the R1200R gratings, respectively. CuNe + CuAr arc lamp spectra were taken just after each target spectrum for wavelength calibration. The spectral templates were only observed with the R1200R grating. In chronological order, the seeing of the WHT spectra was $1.0-1.2$, $0.9-1.1$, $\sim0.8$, $0.7-0.8$, $3.8-3.9$, $1.1-1.3$ and $1.3-1.4$~arcsec.

\subsubsection{Xinglong 2.16-m Telescope}

Four spectra close to orbital phase 0.75 were taken with the 2.16-m telescope at the Xinglong Observatory, China, on 2019 November 14. We used the Beijing-Faint Object Spectrograph and Camera (BFOSC) with the G8 grating and a 1.1-arcsec slit width. The spectral range covered was $6000 -7550$ \AA\ with a dispersion of 1.09 \AA \, $\mathrm{pix}^{-1}$ and a FWHM spectral resolution of 4.8 \AA \, (equivalent to $\simeq 228$ $\mathrm{km}\,\mathrm{s}^{-1}$ at 6300 \AA). We took a spectrum of a  FeAr $+$ Ne arc lamp after each target exposure for wavelength calibration. The seeing varied between $3.0$ and $4.0$~arcsec.

\begin{table}
\caption[]{Log of the spectroscopy observations with the Himalayan Chandra Telescope (HCT), Nordic Optical Telescope (NOT), William Herschel Telescope (WHT) and the 2.16-m telescope at the Xinglong Observatory. The gratings and slit widths (in arcsec) used at the WHT are given in brackets.}
\label{tab:observaciones_espectroscopia}
\centering
\begin{tabular}{l c c c c}
\hline\noalign{\smallskip}
\textbf{Telescope/instrument} &{\#} &$T_\mathrm{exp}$ & Coverage  & $\Delta \lambda$ \\
Date & &(s) & (\AA)  & (\AA) \\
\hline\noalign{\smallskip}
\textbf{HCT/HFOSC} &  &  &\\
2017 Dec 6 & 20 & 600--900  & 5120--9310 & 5.6\\ 
2017 Dec 7 & 24 & 900--1200 & " & "\\ 
2017 Dec 8 & 21 & 1200  & "  & "\\ 
\hline\noalign{\smallskip}
\textbf{NOT/ALFOSC} & & &\\ 
2017 Dec 8 & 30 & 600--900 & 5680--8580  & 3.5\\ 
2017 Dec 9 & 21 & 600 & "  & "\\ 
2017 Dec 10 & 16 & 600 & "  & "\\ 
\hline\noalign{\smallskip}
\textbf{WHT/ISIS} & & &\\ 
2018 Dec 1 (R1200R, 1.0) & 6 & 300 & 5830--6600  & 0.75\\ 
2019 Aug 24 (R600R, 0.7) & 4 & 300 & 5460-6940  & 1.27\\ 
2019 Aug 25 (R600R, 0.7) & 2 & 300 & 5460-6940  & 1.27\\ 
2019 Sep 7 (R1200R, 0.8) & 4 & 300 & 5830--6575  & 0.60\\ 
2019 Sep 8 (R1200R, 1.0) & 4 & 300 & 5830--6575  & 0.75\\ 
2019 Sep 12 (R1200R, 0.8) & 4 & 300 & 5830--6575  & 0.60\\ 
2019 Sep 13 (R1200R, 0.8) & 4 & 300 & 5830--6575  & 0.60\\ 
\hline\noalign{\smallskip}
\textbf{2.16-m Xinglong/BFOSC} & & &\\ 
2019 Nov 14 & 4 & 900 & 6000--7550  & 4.8\\ 
\hline\noalign{\smallskip}
\end{tabular}
\end{table}

\subsection{Photometry}\label{subsec:photometry}
Time-resolved $R$-band photometry was obtained with four telescopes at different geographical longitudes to achieve a good sampling of almost the entire binary orbit. These photometric data were obtained in the period 2017 December--2018 February. Some of the nights were either close in time ($<1$~d) or simultaneous with the NOT and HCT spectroscopy. We used the simultaneous photometry and spectroscopy observations to correct for night-to-night variability in the light curve caused by accretion (Section~\ref{subsec-construction_lc}). The observing log is presented in Table~\ref{tab:observaciones_fotometria}.

\subsubsection{J.\,C. Bhattacharya Telescope}
We obtained time-resolved $R$-band photometry of \targ\ during four nights (2017 December 7--10) using the 1.3-m J.~C. Bhattacharya Telescope (JCBT) in the Vainu Bappu Observatory on the Javadi hills of Tamilnadu, India. This photometry is in part simultaneous with some NOT and HCT spectroscopic data sets (Sections~\ref{subsec-hct} and~\ref{subsec-not}).

We imaged the \targ\ field with the Peltier-cooled Princeton Instruments ProEM CCD camera, an array of $1024 \times 1024$ 13-$\mu\mathrm{m}$ square pixels. This delivers a usable field of view (FOV) of $4.3 \times 4.3$~$\mathrm{arcmin}$ and a pixel size on the sky of 0.26 arcsec. The full-frame, high-gain (5\,MHz frequency) readout mode yielded $1.19\,\mathrm{e}^{-}\,\mathrm{ADU}^{-1}$ and a readout noise of $13\,\mathrm{e}^{-}$. We used the Bessel $R$ filter and fixed the exposure time at 600~s.

\subsubsection{0.4-m University of Athens Observatory}
We used the University of Athens Observatory (UOAO), Greece, 0.4-m robotic and remotely controlled telescope \citep{kosmas-2016} to obtain time-resolved $R$-band photometry of the target on five nights in 2017 December and a further five in 2018 January. Part of the photometry (2017 December 8--10) is simultaneous with some HCT and NOT spectroscopic data sets (Sections~\ref{subsec-hct} and~\ref{subsec-not}). 

We observed \targ\ with the SBIG ST10 CCD camera, an array of $2184 \times 1472$ 6.8-$\mu\mathrm{m}$ square pixels, binned at $2 \times 2$. The FOV was increased to $17 \times 26$~$\mathrm{arcmin}$ with the use of an f/6.3 focal reducer, resulting in a plate scale of $1.39$ $\mathrm{arcsec} \, \mathrm{pixel}^{-1}$. The CCD gain is $1.32\,\mathrm{e}^{-}\,\mathrm{ADU}^{-1}$ and the readout noise $8.8\,\mathrm{e}^{-}$. We used the Johnson-Cousins $R$ filter and an exposure time of 60~s.

\subsubsection{0.3-m Sutter Creek Observatory}
The 0.3-m SC30 telescope located at the Sutter Creek Observatory in California, USA, also provided time-resolved $R$-band photometry of \targ\ on the nights of 2018 January 21 and 28. The observations were carried out with the unbinned $1024 \times 1024$ CCD array of 24-$\mu\mathrm{m}$ pixels. The images cover a $28 \times 28$ arcmin FOV with a plate scale of 1.65 $\mathrm{arcsec} \, \mathrm{pixel}^{-1}$. The CCD readout has a gain of $0.69\,\mathrm{e}^{-}\,\mathrm{ADU}^{-1}$ and a readout noise of 2.81 $\mathrm{e}^{-}$. We used the Johnson-Cousins $R$ filter and the exposure time was fixed at 60~s.

\subsubsection{0.43-m Sierra Remote Observatories}
The 0.43-m f/6.8 CDK telescope at the Sierra Remote Observatories in California, USA, provided further time-resolved $R$-band photometry of \targ\ on 2018 January 25--30 and February 1--3. The target field was imaged on the $2 \times 2$ binned $2004 \times 1336$ CCD array of 18-$\mu\mathrm{m}$ square pixels of the SBIG STXL-11002 camera. This provided a FOV of $42 \times 29$~$\mathrm{arcmin}$ and a plate scale of 1.26 $\mathrm{arcsec} \, \mathrm{pixel}^{-1}$. The readout gain was $1.74\,\mathrm{e}^{-}\,\mathrm{ADU}^{-1}$ and the readout noise 15 $\mathrm{e}^{-}$. We used the Johnson-Cousins $R$ filter with an exposure time of 60~s.

\subsubsection{\textit{TESS} photometry}\label{subsec:TESS}
The Transiting Exoplanet Survey Satellite (\textit{TESS}) is a space-based  optical  telescope  launched in 2018 to perform an all-sky survey to search for transiting exoplanets \citep{ricker-2015}. The telescope consists of four cameras, each with a FOV of $24^{\circ} \times 24^{\circ}$. This results  in  a  combined  telescope FOV of $24^{\circ} \times 96^{\circ}$. The size of each camera is $4096 \times 4096$ pixel and the plate scale is 21 arcsec pixel$^{-1}$. Ninety per cent of the flux of a star is contained within a $4 \times 4$ pixel ($1.4 \times 1.4$ arcmin) region around its centroid \citep{ricker-2015}. \textit{TESS} observations are performed in a single photometric band that covers a broad wavelength range from about 6000 to 11000~\AA.

The satellite observed \targ\ (\textit{TESS} Input Catalog, TIC 431762266) on 2019 November 3--12 and 15--27 (sector 18). The full-frame images of this sector were taken with a cadence of 30 min by combining 900 2-s exposures.

\begin{table}
\caption[]{Log of the time-resolved $R$-band photometry.}
\setlength{\tabcolsep}{0.9ex}
\label{tab:observaciones_fotometria}
\begin{center}
\begin{tabular}{l c c c}
\hline\noalign{\smallskip}
\tabhead{Telescope} &\tabhead{$\#$} &\tabhead{$T_\mathrm{exp}$} & Coverage\\
Date &\tabhead{} &(s) & (h)\\
\hline\noalign{\smallskip}
\textbf{1.3-m JCBT} & &\\
2017 Dec 7 & 37 & 600 & 8.2 \\
2017 Dec 8 & 41 & 600 & 6.9 \\
2017 Dec 9 & 50 & 600 & 8.1 \\
2017 Dec 10 & 46 & 600 &  8.0 \\
\hline\noalign{\smallskip}
\textbf{0.4-m UOAO telescope} &  & \\
2017 Dec 8 & 135 & 60 & 9.9 \\
2017 Dec 9 & 310 & 60 & 9.1 \\
2017 Dec 10 & 594 & 60 & 10.7 \\
2017 Dec 11 & 517 & 60 & 9.3 \\
2017 Dec 01 & 593 & 60 & 11.1 \\
2018 Jan 27 & 400 & 60 & 7.2 \\
2018 Jan 28 & 220 & 60 & 3.9\\
2018 Jan 29 & 385 & 60 & 6.9 \\
2018 Jan 30 & 343 & 60 & 6.2 \\
2018 Jan 31 & 316 & 60 & 6.4 \\
\hline\noalign{\smallskip}
\textbf{0.3-m SC30 telescope} & &\\
2018 Jan 21 & 178 & 60 & 3.6 \\
2018 Jan 28 & 197 & 60 & 4.1 \\
\hline\noalign{\smallskip}
\textbf{0.43-m CDK telescope} & &\\
2018 Jan 25 & 195 & 60 & 5.5 \\
2018 Jan 26 & 230 & 60 & 5.2 \\
2018 Jan 27 & 171 & 60 & 5.5 \\
2018 Jan 28 & 250 & 60 & 5.5 \\
2018 Jan 30 & 173 & 60 & 3.8 \\
2018 Feb 01 & 150 & 60 & 3.4 \\
2018 Feb 02 & 250 & 60 & 5.5 \\
2018 Feb 03 & 100 & 60 & 3.3 \\
\hline\noalign{\smallskip}
\end{tabular}
\end{center}
\end{table}

\subsection{Data reduction}\label{subsec:data_reduction}
All the spectra were reduced, wavelength calibrated and extracted following standard  techniques  implemented in \textsc{iraf}\footnote{{\sc iraf} is distributed by the National Optical Astronomy Observatories.} and \textsc{pamela} \citep[][available in the \textsc{starlink} distribution\footnote{\url{https://starlink.eao.hawaii.edu/starlink}}]{marsh-89}. For the NOT and HCT data the pixel-to-wavelength scale was determined with six-term polynomial fits to 33 and 35 arc lines, respectively. For the WHT and Xinglong telescopes we performed third-order spline fits to 16/23 (R600R/R1200R gratings) and 17 arc lines, respectively. The rms scatter of the fits was $<0.05~$\AA \, for all data sets. We used the  [O\,{\sc i}] 6300.304 \AA\ sky  emission  line  to  look for wavelength zero-point deviations and found they were smaller than the rms scatter of the fitted wavelength calibration, except for the HCT data for which they reach $\simeq10$~km~s$^{-1}$. Hence, we only corrected for these offsets in that case.

The extracted spectra were imported into \textsc{molly}\footnote{\url{http://deneb.astro.warwick.ac.uk/phsaap/software/molly/html/INDEX.html}} in order to do the analysis described in the next sections and corrected for the Earth motion to have them in the heliocentric rest frame. Times are expressed in heliocentric Julian days (UTC). Finally, they were normalised using a seventh-order polynomial fit to the continuum after masking the strong emission lines.

The $R$-band images were debiased and flat-fielded using the standard CCD data processing workflow within {\sc iraf}. Differential photometry with variable aperture was performed with the HiPERCAM pipeline\footnote{\url{https://github.com/HiPERCAM}}. For this purpose, we used the field stars GK Per--12 ($r \simeq 13.4$~mag) and GK Per--10 ($r \simeq 14.4$~mag) labelled in \cite{hh-95,hh-97} as the comparison and check star, respectively.

We obtained the \textit{TESS} light curve from the 30-min full-frame images using the {\sc tesseract}\footnote{\url{https://github.com/astrofelipe/tesseract##readme}} package (Rojas et al., in prep.) that performs aperture photometry via {\sc TESSCut} \citep{brasseur-19} and {\sc lightkurve} \citep{lightkurve-18}. Visual inspection of the \targ\ field revealed some contaminating stars given the limited angular resolution of the \textit{TESS} images. Using the \textit{Aladin Sky Atlas}\footnote{\url{https://aladin.u-strasbg.fr/}} and the Pan--STARRS Data Release 1 \citep{chambers-2016} we checked that the six brightest objects in the $4 \times 4$ pixel region around \targ\ have $r=14.7-17.8$~mag, fainter than \targ\ ($r \simeq 12.7$~mag) and non-variable. Hence, these contaminating stars only add a constant veiling to the \targ\ light curve. We used an on-target photometric aperture of one pixel in order to minimise this contamination. A larger circular aperture was checked to result in a significant decrease of the light curve amplitude.

\section{Analysis and results}
All uncertainties presented in this and the next sections are quoted at $68$ per cent confidence unless otherwise stated.

\begin{figure*}
\centering \includegraphics[height=10cm]{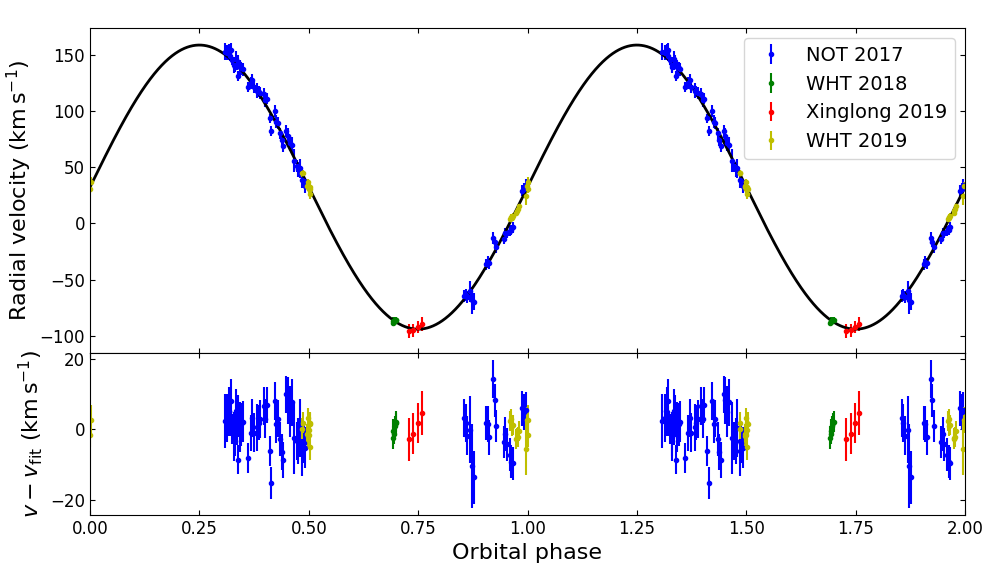}
\caption{\label{fig:radialvelocitycurve} Top panel: Heliocentric radial velocity curve of the donor star absorption features obtained by cross-correlating the individual spectra with the spectral template HD~20165 (K1~V). The error bars have been scaled by a factor $\simeq 1.26$ to obtain a fit with $\chi^{2}/\mathrm{dof} = 1.0$. The best sine fit is shown as a black line. The orbital cycle has been repeated for the sake of clarity. Bottom panel: residuals of the fit.}
\end{figure*}

\begin{table*}
\caption[]{Radial velocity curve best-fit parameters. The rms of the wavelength calibration was lineally added to the uncertainties of the $\gamma$ values. Degrees of freedom ($\mathrm{dof}) =94$.}
\centering
\begin{tabular}{lccccc}
\hline\noalign{\smallskip}
Template & Spectral type & $\gamma$ & $K_2$ & $P$ & $T_0$\\
 &  & (\kms) & (\kms) & (d) & (HJD) \\
\hline\noalign{\smallskip}
HD 20165 & K1 V & $32 \pm 2$ & $126.4 \pm 0.9$  & $1.996872 \pm 0.000009$ & $2458095.709 \pm 0.002$\\
HR 2556 & K0 III-IV & $33 \pm 2$ & $126.9 \pm 0.9$  & $1.996874 \pm 0.000009$ & $2458095.709 \pm 0.003$\\
 \hline\noalign{\smallskip}
\end{tabular}
\label{tab:rvcurve_params}
\end{table*}

\subsection{Radial velocity curve of the donor star}\label{sec-k2}
We measured the radial velocities of the donor absorption lines by cross-correlating each \targ\ spectrum with the spectrum of the K1\,V HD~20165 template star in the spectral range $6050-6538$~\AA, after masking the diffuse interstellar band at $\approx 6280$~\AA. Prior to this, the template spectrum was corrected for its systemic velocity and for any wavelength zero-point offset by removing the velocity measured by Gaussian fitting the core of the H${\alpha}$ absorption line. Also, all the spectra were re-binned on to a common constant velocity scale. We proceeded in the same way with the K0~III-IV HR~2556 template, which resulted in very similar values of the radial velocities. To account for all uncertainties, the rms of the wavelength calibration was added linearly to the statistical uncertainty of each radial velocity measurement. Since there is no evidence for irradiation of the donor star (see Section~\ref{sec-vsini}), we performed least-squares sinusoidal fits to the radial velocities, $V (t)$, of the form:

\begin{equation}
\label{eq:seno}
V (t) = \gamma + K_{2} \, \mathrm{sin} \, \left[ \frac{2 \pi}{P}(t-T_{0}) \right] ,
\end{equation}  

\noindent where $\gamma$ is  the  heliocentric systemic velocity, $K_2$ the  radial  velocity semi-amplitude of the donor star, $P$ the orbital period and $T_0$ the time of closest approach of the donor star to the observer. Fig.~\ref{fig:radialvelocitycurve} shows the radial velocity curve and the best fit. In our preliminary fits, the HCT/HFOSC radial velocities showed a large scatter that the rest of data did not at similar orbital phases, with a deviation of up to 45-$\sigma$ from the initial best fit. We could not identify the reason for this and we excluded these data from the fitting process. In addition, one NOT/ALFOSC radial velocity point was discarded since its deviation was larger than 7-$\sigma$. After rejecting these deviant points, the $\chi^{2}$ relative to the number of degrees of freedom (dof),  $\chi^{2}/\mathrm{dof}$, was $\simeq 1.6$ for both templates. We followed by rescaling the radial velocity uncertainties by a factor $\simeq 1.26$ so that the $\chi^{2}/\mathrm{dof}$ of the fit was $1.0$. The best-fit parameters for each template are listed in Table~\ref{tab:rvcurve_params}. 

We obtained an orbital period that agrees within 1-$\sigma$ of the value $P=1.9968 \pm 0.0008\,\mathrm{d}$ reported in MR02. Our $K_2$ determination agrees within 1-$\sigma$ of the value $K_2=124 \pm 2$\,\kms\ reported by \cite{crampton-1986}. Their coverage of the radial velocity curve only showed small gaps around orbital phases $0.0$, $0.35$, $0.45$ and $0.9$, and they achieved a good sampling of the orbit quadratures\footnote{Note that \cite{crampton-1986} defined $T_0$ as the time of maximum radial velocity of the donor star and here and in Section~\ref{sec-spectralclassification} we provide their orbital phase coverage according to our $T_0$ convention.}. On the other hand, our $K_2$ is only consistent  with $K_2=120.5 \pm 0.7$\,\kms\ obtained by MR02 at the 4-$\sigma$ level. They cover the orbital phases $0.0$ to $0.3$ and $0.5$ to $0.8$. However, their radial velocity curve displays a significant number of departing points at phases $0.25-0.30$, which probably acted to lower the amplitude of their best sine fit (see fig.~3 in their paper). These authors also obtained a value of $K_2=129 \pm 2.7$\,\kms\ (consistent with ours at the 1-$\sigma$ level) by combining the radial velocities measured by \cite{kraft-64}, \cite{crampton-1986} and \cite{reinsch-94}. Finally, \cite{crampton-1986} and MR02 obtained a systemic velocity of $28 \pm 1$ and $40.8 \pm 0.7$\,\kms, respectively. Our best-fit $\gamma \simeq 32$~km~s$^{-1}$ lies between those two values, but is much closer to the Crampton et al.’s estimate.

\subsection{Spectral classification of the donor star}\label{sec-spectralclassification}
The orbital period-mean density ($\rho_2$) relation for Roche lobe-filling stars \citep{faulkneretal72} yields $\rho_2 \, (\mathrm{g/cm^{3}}) \simeq 110 \, P (\mathrm{h})^{-2} = 0.048 \, \mathrm{g/cm^{3}} = 0.034 \rho_{\odot}$ for the evolved donor star in \targ. This value is in between those expected for main sequence and giant stars (e.g. $\simeq 1.26\,\rho_{\odot}$ and $\simeq 0.00031\,\rho_{\odot}$ for K0~V and K0~III, respectively, \citealt{cox-2000}). In order to constrain the spectral type of the donor star we used two grids of high resolution ($R = \lambda / \Delta \lambda \simeq 60000$) templates covering $4990 - 6410$~\AA\ and extracted from the library published in \cite{yee-2017}. The first grid (Table~\ref{tab:spectraltype-ms}) contains nine spectra of main sequence stars ($\mathrm{log} \, g = 4.4-4.6$~dex) in the range G3~V--K4~V with $-0.10$~dex~<~$\mathrm{[Fe/H]} < 0.10$~dex metallicity. The spectral type of each template was determined according to its effective temperature \citep{yee-2017} and the canonical value for each spectral type \citep{pecaut-13}. The second grid (Table~\ref{tab:spectraltype-sg}) includes eight spectra of subgiant stars with $\mathrm{log} \, g = 3.0-3.6$~dex, which is close to the surface gravity of the donor star in \targ\ (Section~\ref{discussion-masses}). This grid covers effective temperatures from $\simeq 4800$ to $\simeq 5300 \, \mathrm{K}$ and its metallicity is also $-0.10$~dex~<~$\mathrm{[Fe/H]} < 0.10$~dex.

We used the WHT/ISIS R1200R spectra taken on 2019 September 12 and 13 that cover the orbital phases $0.5$ and $0.96$, respectively. We selected these data sets to search for potential phase-dependent changes in the spectral classification due to irradiation of the inner face of the donor star by the WD and/or accretion structures. The templates were downgraded to match the resolution of the \targ\ spectra by convolution with Gaussian profiles. Then, we applied the optimal subtraction technique described in \cite{marsh94} to every template. We performed this analysis in the spectral range $6050-6400$~\AA. We proceeded as follows: we corrected for the radial velocity of each \targ\ spectrum to velocity-shift them to the rest frame of the template. Next, we computed a weighted average of the \targ\ spectra giving larger weights to those with higher signal-to-noise ratio. We subsequently broadened the photospheric lines of the template spectra by convolution with the Gray's rotational profile \citep{libro-gray}, probing the $v_\mathrm{rot} \sin i$ space between $1$ and $150$\,\kms\ in steps  of  $1$\,\kms. A robust measurement of $v_\mathrm{rot} \sin  i$ will be given in Section~\ref{sec-vsini}, where we will use templates taken with the same instrumental setup as the \targ\ data. We used a linear limb-darkening coefficient of $0.65$, a reasonable choice for a K0--3~IV star \citep{claret95}. Note that similar results are obtained for values of $0.5$ or $0.8$. The broadened versions of each template spectrum were multiplied by a factor $f$ between $0$ and $1$ and then subtracted from the weighted-average spectrum of \targ. This factor represents the fractional contribution of the donor star to the total flux in the wavelength range of the analysis. Finally, we searched for the values of $v_\mathrm{rot} \sin  i$ and $f$ that minimised the $\chi^2$ between the residual of the subtraction and a smoothed version of itself obtained by convolution with a 15-\AA \, FWHM Gaussian. In doing so, we compared the results of using  different FWHMs (between $15$ and $40$~\AA) for the smoothing Gaussian. The results were found to be the same within the uncertainties. The minimum $\chi^{2}/\mathrm{dof}$ for each template is presented in Tables~\ref{tab:spectraltype-ms} and~\ref{tab:spectraltype-sg}. 

The results obtained with the main sequence templates suggest a spectral type of the donor star in the range G7--K1 with the lowest $\chi^{2}/\mathrm{dof}$ value found for K1. On the other hand, the $\chi^{2}/\mathrm{dof}$ values obtained with the grid of subgiant templates provide an effective temperature in the range $\simeq 4900-5150$~K. In this regard, \citet[][see also \citealp{harrison-2015}]{harrison-16} characterised the chemical composition of the donor star in \targ\ using near-infrared spectroscopy and synthetic spectral templates with surface gravity $\log\,g = 4.0$~dex. The effective temperature of their best-fit template was $5000 \, \mathrm{K}$, with an estimated uncertainty of $\pm 110 \, \mathrm{K}$, in agreement with our constraint.

\cite{kraft-64} and \cite{gallagher-74} noticed potential spectral type changes with orbital phase. However, \cite{crampton-1986} found that the average spectra at phases $0.55$ and $0.05$ (when we observe the hemisphere of the donor facing and trailing the WD, respectively) were consistent with being identical. Similarly, MR02 found no differences in the spectral type of the donor star between the phase intervals $0.03-0.28$ and $0.54-0.82$ when applying the optimal subtraction method to their $B$-band spectra. Our analysis also showed no noticeable changes between phases $0.5$ and $0.96$. We thus conclude that UV and X-ray heating of the donor star is most likely negligible during the quiescence state.

\begin{table}
\caption[]{Spectral classification of the donor star in \targ\ using WHT/ISIS data and spectra of main sequence stars as templates. The number of $\mathrm{dof}$ is $1170$.}
\centering
\begin{tabular}{lcccccc}
\hline\noalign{\smallskip}
Template & Spectral  & Effective & \multicolumn{2}{c}{$\chi^2_{\mathrm{min}}/\mathrm{dof}$ at} \\
\tabhead{} & type  & temperature & \multicolumn{2}{c}{orbital phase:}  \\
\tabhead{} & \tabhead{} & (K)  & $0.5$ & $0.96$ \\
\hline\noalign{\smallskip}
HD 42807 & G3 V & $5730 \pm 60$ & 1.53 & 1.66 \\
HD 43162 & G6 V & $5617 \pm 60$ & 1.54 & 1.64 \\
HD 10780 & G9 V & $5398 \pm 75$ & 1.41 & 1.48 \\
HD 72760 & K0 V & $5293 \pm 60$ & 1.42 & 1.45 \\
HD 110743 & K1 V & $5198 \pm 60$ & 1.36 & 1.38 \\
HD 8553 & K2 V & $5053 \pm 60$ & 1.53 & 1.52 \\
HD 153525 & K3 V & $4826 \pm 60$ & 1.68 & 1.66\\
HIP 118261 & K4 V & $4615 \pm 214$ & 1.89 & 1.87\\
 \hline\noalign{\smallskip}
\end{tabular}
\label{tab:spectraltype-ms}
\end{table}

\begin{table}
\caption[]{Spectral classification of the donor star in \targ\ using WHT/ISIS data and spectra of subgiant stars as templates. The number of $\mathrm{dof}$ is $1170$.}
\centering
\begin{tabular}{lccccc}
\hline\noalign{\smallskip}
Template & Effective  & \multicolumn{2}{c}{$\chi^2_{\mathrm{min}}/\mathrm{dof}$ at} \\
\tabhead{} & temperature  & \multicolumn{2}{c}{orbital phase:}  \\
\tabhead{} & (K)  & $0.5$ & $0.96$ \\
\hline\noalign{\smallskip}
HD 77818 & $4777 \pm 60$ & 1.47 & 1.35 \\
HD 95526 & $4832 \pm 60$ & 1.36 & 1.28 \\
HD 31451 & $4921 \pm 60$ & 1.27 & 1.20 \\
HD 40537 & $4962 \pm 60$ & 1.33 & 1.24 \\
HD 122253 & $5007 \pm 60$ & 1.24 & 1.19 \\
HD 14855 & $5023 \pm 60$ & 1.24 & 1.15 \\
HD 17620 & $5159 \pm 60$ & 1.24 & 1.21 \\
HD 108189 & $5373\pm 60$ & 1.35 & 1.40\\
 \hline\noalign{\smallskip}
\end{tabular}
\label{tab:spectraltype-sg}
\end{table}

\subsection{Binary mass ratio} \label{sec-vsini}
The donor-to-WD mass ratio ($q=M_2/M_1$) is related to $K_2$ and $v_\mathrm{rot} \sin  i $ through:
\begin{equation}
\label{eq:vsini}
v_\mathrm{rot} \sin  i  \simeq 0.49 (1+q) q^{2/3}  K_2 \left[0.6 q^{2/3} + \mathrm{ln} (1+q^{1/3})  \right]^{-1} \, .
\end{equation}

\begin{figure}
\centering \includegraphics[height=5cm]{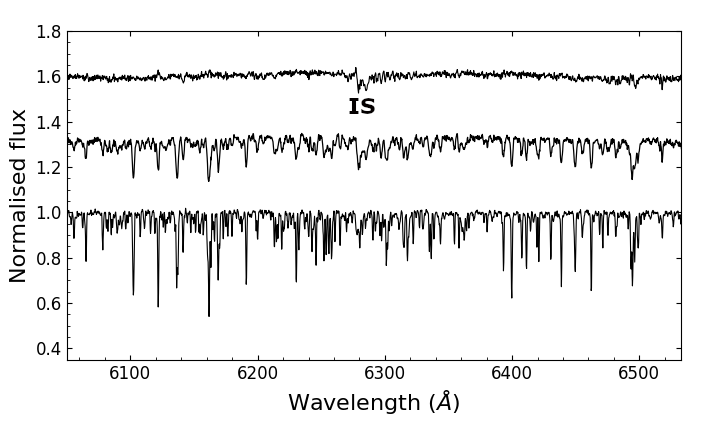}
\caption{\label{fig:optsub} Result of the optimal subtraction technique to measure $v_\mathrm{rot} \sin  i$ and $f$ from the WHT/ISIS spectra taken on 2018 December 1. From bottom to top: spectrum of the K1~V template HD~20165, the average spectrum of \targ\ in the rest frame of the donor star and the residual spectrum after subtraction of the broadened and scaled template. The spectra have been shifted vertically for display purposes. "IS" marks an interstellar absorption band contaminated with telluric absorption.}
\end{figure}

\noindent
This relation is obtained adopting the Eggleton's approximation for the Roche lobe radius \citep{eggleton-83} and under the assumptions that the orbit of the system is circular, the angular momentum vector of both the orbit and the donor star are aligned and that their rotation is synchronized as a result of tidal interactions. Hence, $q$ can be derived from $K_2$ and $v_\mathrm{rot} \sin  i$. The latter is provided by the subtraction of stellar templates described in Section~\ref{sec-spectralclassification}. Here we apply this technique to all our spectra in the same wavelength range as used in the radial velocity analysis (Section~\ref{sec-k2}) with the HD~20165 and HR~2556 spectral templates. The templates were observed with the same instrumental setup as the target, except for the WHT/ISIS R600R data. In this case, we used the WHT/ISIS R1200R templates smoothed with a Gaussian profile to match the spectral resolution. Fig.~\ref{fig:optsub} displays the normalised, Doppler-shifted average of \targ\ before and after the subtraction of the broadened K1~V template.

Evaluation of the uncertainties for $v_\mathrm{rot} \sin  i$ and $f$ was performed by Monte Carlo randomization following the approach in \cite{steeghs-2007} and \cite{torres-2020}. The optimal subtraction procedure was repeated for 10000 bootstrapped copies of the \targ\ average spectrum. This delivered the distributions of possible values for $v_\mathrm{rot} \sin  i$ and $f$, which are well fitted by Gaussians. Hence, we took their mean and standard deviation as the value and 1-$\sigma$ uncertainty, respectively (Table~\ref{tab:vsini_f_hd20165}). The rotational velocities obtained from the NOT, HCT and Xinglong spectra are overestimated and/or have large uncertainties as a result of the lower spectral resolution ($\simeq 170-270$\,\kms; see Table~\ref{tab:observaciones_espectroscopia}). The spectral resolution of the WHT/ISIS R600R data is comparable to the $v_\mathrm{rot} \sin  i$ of the system, but the templates were not obtained with the same instrumental setup. For these reasons, only the measurements of $v_\mathrm{rot} \sin  i$ from the WHT/ISIS R1200R data will be considered.

The observed variability of $v_\mathrm{rot} \sin  i$ with the orbital phase (see Table~\ref{tab:vsini_f_hd20165}) may be compatible with that expected for a Roche lobe-filling donor star \citep[see e.g.][]{shahbaz-14}. However, our sampling is insufficient to establish the phase dependence of $v_\mathrm{rot} \sin  i$: only critical orbital phases were covered to estimate its mean value. Taking the averages at phases $\approx 0.0$, 0.5 and 0.7 we derive $v_\mathrm{rot} \sin  i = 51 \pm 2 \, \mathrm{km~s^{-1}}$ and  $v_\mathrm{rot} \sin  i = 54 \pm 2 \, \mathrm{km~s^{-1}}$ for HD~20165 and HR~2556,  respectively. Given that the minimum $\chi^2$ value of the optimal subtraction is similar for both templates, we adopt the mean $v_\mathrm{rot} \sin  i = 52 \pm 2$\,\kms. Our result is a significant improvement on the previous estimates of $61.5 \pm 11.8  \, \mathrm{km~s^{-1}}$ (MR02) and $55 \pm 10$\,\kms\ \citep{harrison-2015}.

We derived $q$ using  Eq.~\ref{eq:vsini} and $v_\mathrm{rot} \sin  i$ and $K_2$ (Section~\ref{sec-k2}). To compute its uncertainty, we followed a Monte Carlo approach: we picked random values of $K_2$ and $v_\mathrm{rot} \sin  i$ from normal distributions defined by the mean and the 1-$\sigma$ uncertainties of our measurements. We then calculated $q$ for each random set of parameters and repeated this process 10000 times. The resulting values of $q$ also followed a normal distribution and hence we took the mean and the standard deviation as reliable estimates of its value and uncertainty, respectively. We finally obtain a binary mass ratio:
$$q=0.38 \pm 0.03\,.$$
\noindent MR02 reported a highly uncertain $q=0.55 \pm 0.21$ using the same technique with lower resolution $B$-band spectra. \cite{crampton-1986} estimated $q=0.28 \pm 0.04$ from the quotient of the donor and the H$\beta$ radial velocity semi-amplitudes. This discrepancy indicates that the H$\beta$ emission line is indeed not a good tracer of the WD motion, as pointed out by MR02.\\


\begin{table}
\caption[]{$v_\mathrm{rot} \sin  i$ and $f$ from the optimal subtraction of the HD~20165 (K1 V) and HR~2556 (K0 III-IV, in brackets) template spectra. The $^*$ marks the WHT/ISIS R1200R observations, that provided robust measurements of $v_\mathrm{rot} \sin  i$.}
\label{tab:vsini_f_hd20165}
\centering
\begin{tabular}{l c c c}
\hline\noalign{\smallskip}
\tabhead{Telescope} &Mean & $v_\mathrm{rot} \sin  i$ &$f$\\
Date &orbital phase & (\kms) &\\ \hline\noalign{\smallskip}
\textbf{HCT} &  &  & \\
2017 Dec 06 & 0.26 & $72 \pm 10$ & $0.67 \pm 0.02$\\
& & ($85 \pm 10$) & ($0.73 \pm 0.02$)\\
& & & \\
2017 Dec 07 & 0.76 & $72 \pm 8$ & $0.69 \pm 0.02$\\
& & ($85 \pm 9$) & ($0.75 \pm 0.02$)\\
& & &\\
2017 Dec 08 & 0.26 & $64 \pm 9$ & $0.69 \pm 0.02$\\
& & ($81 \pm 8$) & ($0.75 \pm 0.02$) \\
\hline\noalign{\smallskip}
\textbf{NOT} &\\
2017 Dec 08 & 0.39 & $65 \pm 5$ & $0.66 \pm 0.01$ \\
& & ($72 \pm 5$) & ($0.74 \pm 0.02$) \\
& & &\\
2017 Dec 09 & 0.92 & $63 \pm 5$ & $0.69 \pm 0.02$\\
& & ($65 \pm 5$) & ($0.79 \pm 0.01$)\\
& & &\\
2017 Dec 10 & 0.40 & $66 \pm 5$ & $0.70 \pm 0.01$\\
& & ($76 \pm 5$) & ($0.79 \pm 0.02$)\\
\hline\noalign{\smallskip}
\textbf{WHT} &\\
2018 Dec 01* & 0.70 & $53.8 \pm 0.7$ & $0.71 \pm 0.01$ \\
& & ($56.1 \pm 0.8$) & ($0.77 \pm 0.01$)\\
& & &\\
2019 Aug 24 & 0.00 & $48 \pm 3$ & $0.61 \pm 0.02$\\
& & ($53 \pm 3$) & ($0.70 \pm 0.02$)\\
& & &\\
2019 Aug 25 & 0.48 & $45 \pm 2$ & $0.61 \pm 0.01$\\
& & ($48 \pm 2$) & ($0.70 \pm 0.01$)\\
& & &\\
2019 Sep 7* & 0.98 & $48.5 \pm 0.5$ & $0.71 \pm 0.01$\\
& & ($50.8 \pm 0.6$) & ($0.78 \pm 0.01$)\\
& & &\\
2019 Sep 8* & 0.50 & $50.5 \pm 0.8$ & $0.67 \pm 0.01$\\
& & ($51.7 \pm 0.9$) & ($0.72 \pm 0.02$)\\
& & &\\
2019 Sep 12* & 0.50 & $52.6 \pm 0.5$ & $0.59 \pm 0.01$\\
& & ($53.8 \pm 0.5$) & ($0.65 \pm 0.01$)\\
& & &\\
2019 Sep 13* & 0.96 & $48.1 \pm 0.6$ & $0.65 \pm 0.01$\\
& & ($50.7 \pm 0.6$) & ($0.71 \pm 0.01$)\\
\hline\noalign{\smallskip}
\textbf{2.16m-Xinglong} &\\
2019 Nov 14 & 0.74 & $79 \pm 7$ & $0.60 \pm 0.01$\\
& & ($82 \pm 8$) & ($0.64 \pm 0.02$)\\
\hline\noalign{\smallskip}
\end{tabular}
\end{table}

\subsection{Ellipsoidal light curve and orbital inclination}\label{sec-lightcurve}

In this section we model the $R$-band light curve of \targ. We start by detailing the steps followed to obtain an ellipsoidal light curve as free as possible from night-to-night variations due to accretion. Then, we present and discuss the light curve modelling and provide the binary inclination.

\begin{figure*}
\centering \includegraphics[height=14cm]{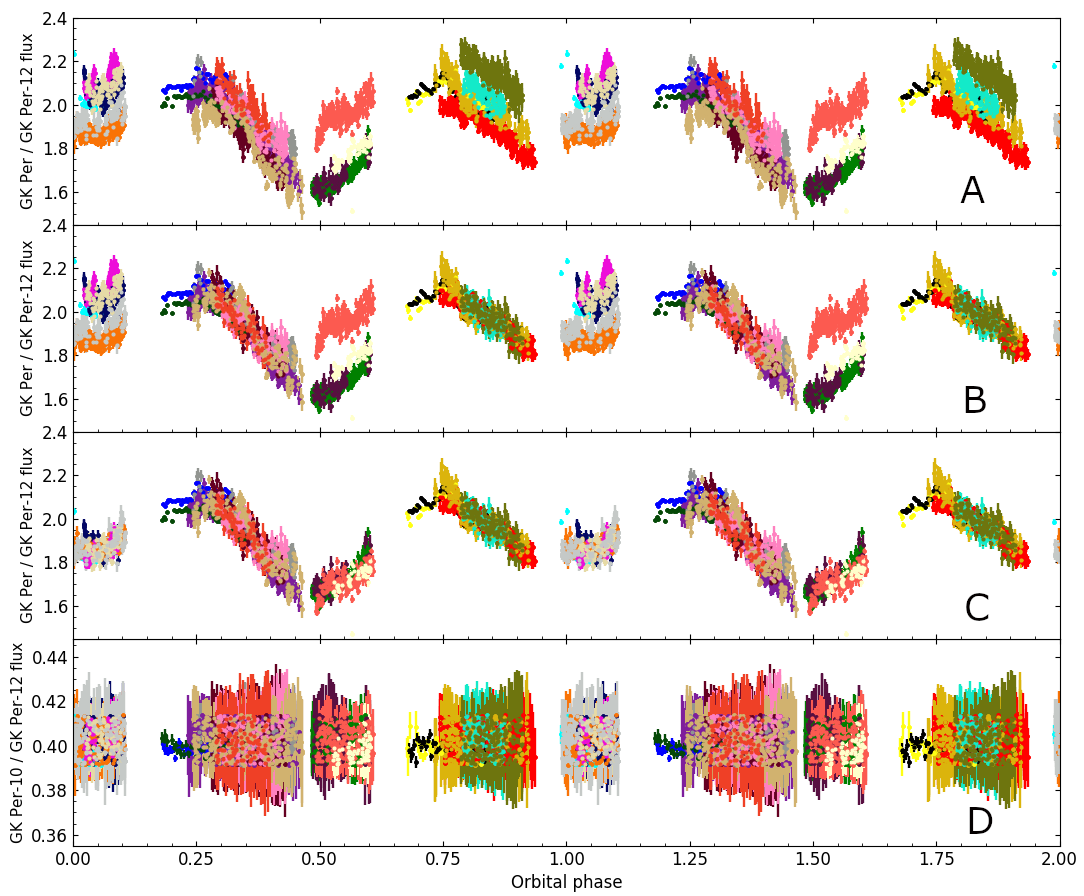}
\caption{\label{fig:steps_lc_correction} Construction of the phase-folded $R$-band light curve of \targ. Panels A to C illustrate the steps followed to correct the photometry for night-to-night accretion variability, while panel D displays the flux of the GK Per-10 check star relative to the GK Per-12 comparison star. Panel A shows the light curve prior to applying any variability correction. Panel B displays the light curve after shifting the relative flux of the data points that cover similar orbital phases than the 2017 December 7--10 photometry. The shifts are applied in order to match the mean flux measured during those above nights. Panel C shows the result after correcting the flux points at other orbital phases. Each colour in the plot represents a different night. See text for more detail. Two cycles are shown for the sake of clarity.}
\end{figure*}

\subsubsection{Multi-epoch $R$-band light curve}\label{subsec-construction_lc}
We constructed the phase-folded $R$-band light curve of \targ\ using the photometry data described in Section~\ref{subsec:photometry} (see Table~\ref{tab:observaciones_fotometria}) and the ephemeris obtained in Section~\ref{sec-k2}. In order to exclude the data points affected by large systematic errors we examined the flux stability of the GK Per-10 check star relative to the GK Per-12 comparison star (see Section~\ref{subsec:data_reduction}). After some testing, we removed the points with a relative deviation $>0.03$\,mag from the mean. Similarly, points with a statistical uncertainty $>0.03$\,mag were also removed. The bottom panel of Fig.~\ref{fig:steps_lc_correction} shows the final GK Per-10/GK Per-12 relative flux curve.

The cleaned, phase-folded $R$-band light curve of \targ\ (top panel of Fig.~\ref{fig:steps_lc_correction}) shows clear night-to-night variations likely due to changes in the light contribution of the accretion flow. To correct for these, we took advantage of the partially simultaneous photometry and spectroscopy on 2017 December 7--10. The photometry yielded flux points that cover the orbital phases $0.2-0.5$ and $0.65-0.95$, while the spectra provided a nearly constant $f$ for those nights (see Table~\ref{tab:vsini_f_hd20165}). The correction consisted of adding or subtracting a constant value to shift the photometry data of all the other nights to match the flux of the above four reference nights. This was accomplished in three steps: first, the light curves that cover the same orbital phases as the reference nights were shifted to the mean reference flux level in the range of coincidence. The resulting light curve is shown in panel B of Fig.~\ref{fig:steps_lc_correction}. Second, we applied the flux shifts obtained in the previous step to the light curves that sample different orbital phases during the same nights. By doing this, we are assuming that the variability is negligible for time intervals shorter than one day. Finally, the remaining observations were offset to match the mean relative flux of the data points corrected in the second step. The resulting $R$-band light curve is shown in panel C of Fig.~\ref{fig:steps_lc_correction}. Given the steps followed above, the value of the fractional contribution of the donor star to the total flux in the light curve should match that obtained from the spectroscopy on the reference nights ($f=0.66-0.70$ or $0.73-0.79$, depending on the template; see Table~\ref{tab:vsini_f_hd20165}). After the above corrections, the ellipsoidal modulation becomes apparent in the light curve with a peak-to-peak amplitude of $\approx 0.25$\,mag.

We also constructed the phase-folded \textit{TESS} light curve (Section~\ref{subsec:TESS}) using our ephemeris. We only combined six out of a total of eight full orbital cycles since the system appeared to be increasing in brightness during the first two. Prior to phase-folding, we flux-shifted the data corresponding to individual full orbital cycles in order to have all of them with a common mean flux level.

\begin{figure*}
\centering \includegraphics[height=7.5cm]{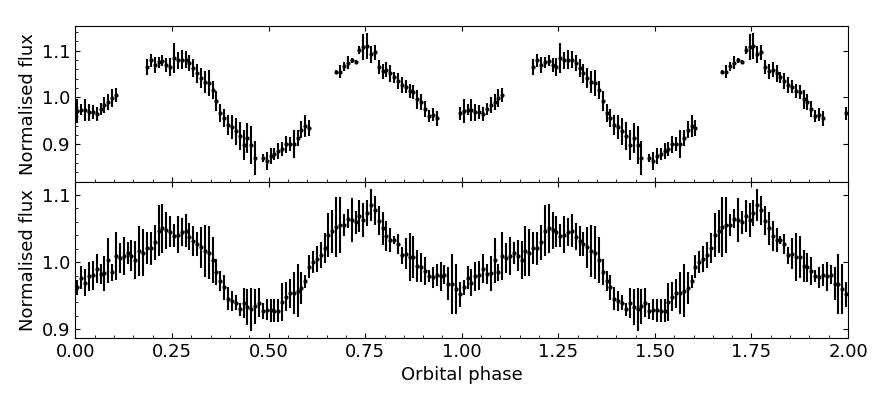}
\caption{\label{fig:lc_r_y_tess.png} Top panel: Phase-folded $R$-band light curve of \targ\ after correcting for night-to-night accretion variability. Bottom panel: \textit{TESS} light curve obtained by phase folding six orbital cycles scaled to have the same mean flux. A 0.01-phase binning has been applied to both data sets.}
\end{figure*}

Figure~\ref{fig:lc_r_y_tess.png} shows the $R$-band (top panel) and the \textit{TESS} (bottom panel) light curves of \targ\ after applying a $0.01$-phase binning and dividing them by their mean flux. While the former shows maxima consistent with being identical in amplitude, the \textit{TESS} light curve hints to unequal maxima. Hence multi-colour photometry with good coverage of the full orbit will allow to check for this behaviour, which might be related to a disc hot spot and/or a spotted donor star. On the other hand, the phase-folded $R$-band photometry deviates from the expected ellipsoidal modulation at phases $0.0-0.1$ and the slope of the individual $R$-band light curves in that phase range shows significant night-to-night variation (see panels A to C of Fig.~\ref{fig:steps_lc_correction}). For this reason we decided to exclude the $0.0-0.1$ orbital phase range from the modelling.

\subsubsection{Light curve modelling}\label{subsec-lcm}
We modelled the $R$-band light curve by fitting synthetic light curves generated with {\sc XRBinary}, a code developed by E.~L. Robinson\footnote{A detailed description of {\sc XRBinary} can be found at: \url{http://www.as.utexas.edu/~elr/Robinson/XRbinary.pdf}}. It accounts for the photometric modulation of a binary system composed of a primary star (assumed to be a point source) surrounded by an accretion disc and a co-rotating Roche lobe-(fully) filling donor star. The disc can be non-axisymmetric and vertically extended. The code also allows for an outer disc rim, an inner torus and disc spots of different brightness. The flux spectrum of the donor star is computed from the stellar atmosphere models of \cite{kurucz-96} using a non-linear limb-darkening law \citep{claret-2000-ld} valid for $\log \, g = 0.0 - 5.0$~dex. The gravity darkening only depends on the star's effective temperature and is based on \cite{claret-2000-gd}. The accretion disc is assumed to be optically thick and to emit as a multi-temperature blackbody. The disc temperature radial profile is given by $T=\mathrm{K}\,r^{\beta}$, where $\mathrm{K}$ is a normalisation constant and $r$ is the distance to the primary star (see the {\sc XRBinary} manual for further details). Other accretion structures and the primary star are also assumed to emit as black bodies. Ray tracing is used to compute the light curve, that can be generated for the Johnson-Cousins filters or for square bandpasses \citep{bayless-2010}.

The disc opening angle can be estimated as
$\alpha \simeq 2 \, \arctan [0.038 (\dot{M}/10^{16})^{3/20}]$ \citep{warner-libro}. Following \cite{webbink-87} and \cite{anupama-93} we calculated an accretion rate in the disc $\dot{M}=(1.0-3.8) \times 10^{17} \, \mathrm{g} \, \mathrm{s}^{-1}$ for orbital inclinations and WD masses in the ranges $55^{\circ}-72^{\circ}$ and $0.8-1.2 \, \mathrm{M}_{\odot}$, respectively. On the other hand, \cite{bianchini-83} and \cite{wada-2018} presented values very close to $\dot{M} \simeq 10^{16} \, \mathrm{g} \, \mathrm{s}^{-1}$. Considering the accretion rate to be in the range  $\dot{M} \simeq 10^{16} - 10^{17} \, \mathrm{g}~ \mathrm{s}^{-1}$, we obtain $\alpha \approx 4^{\circ}-6^{\circ}$. For the light curve modelling we adopted the upper limit, although using a flat disc ($\alpha \simeq 0^{\circ}$) produces the same results. We also assumed a circular disc extending up to the circularisation radius $R_\mathrm{c}=(1+q) \, (b_1/a)^{4}$, where $b_1$ is the distance from the primary star to the inner Lagrangian point of the system, with $b_1/a = (1.0015+q^{0.4056})^{-1}$ \citep{warner-libro}. 

We fixed the donor star effective temperature ($T_2$) at $5000$~K following the spectroscopic measurement by \cite{harrison-16}, which is supported by our analysis in Section~\ref{sec-spectralclassification}. We did not include either a disc hot spot or donor star spots given that the $R$-band light curve has its two maxima at the same flux level within the errors. In addition, we could not place constrains on the temperature of the disc outer edge or the albedos of the donor star and the disc. However, we checked that these parameters have a negligible impact on the light curve modelling and we fixed them at $2000 \, \mathrm{K}$ and $0.5$, respectively. The free parameters in our model are the orbital inclination ($i$), the bolometric luminosity of the disc ($L_\mathrm{D}$), the exponent of the disc temperature radial profile ($\beta$), the disc inner radius  ($R_{\mathrm{in}}$), $q$ and $K_2$. 

\begin{figure*}
\centering \includegraphics[height=18cm]{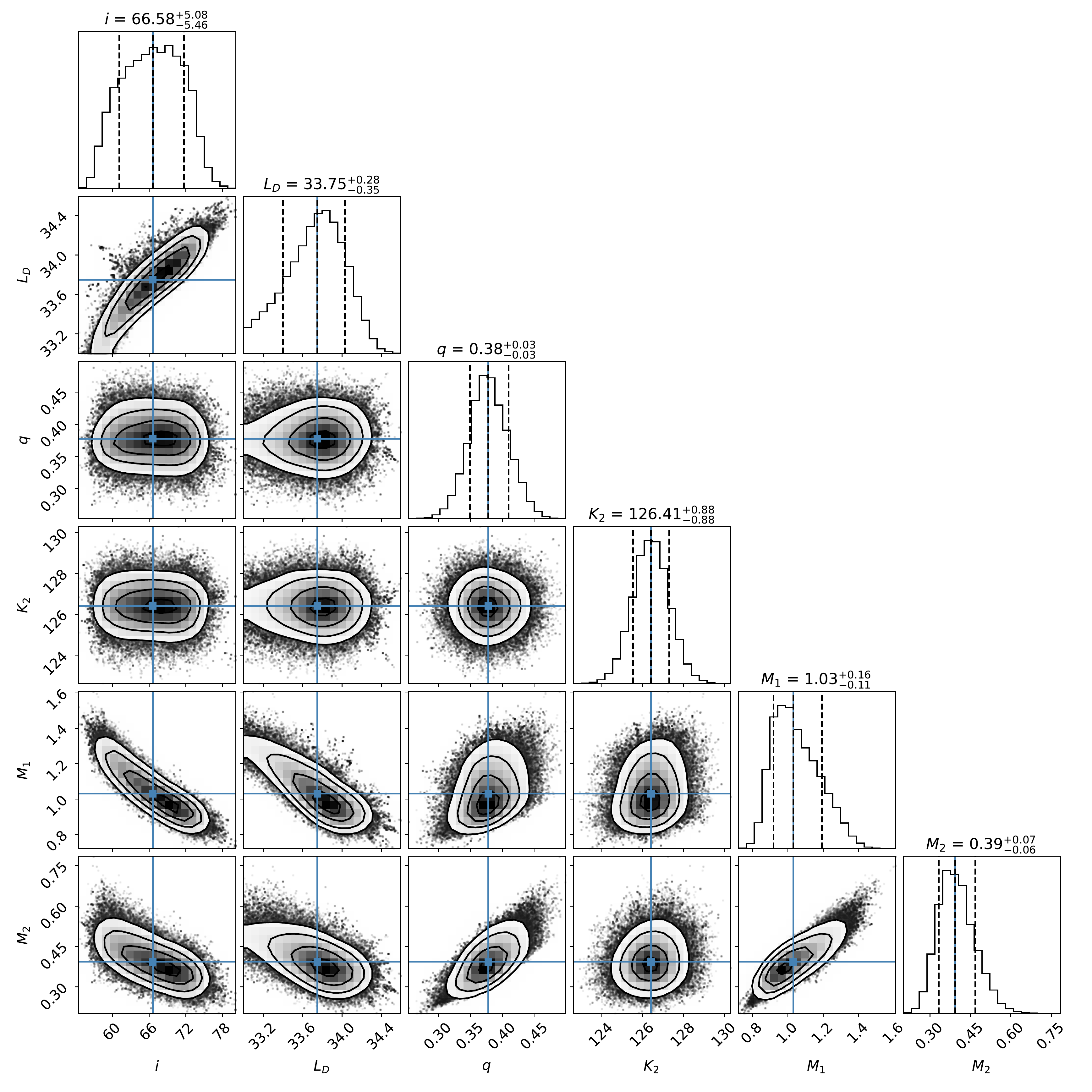}
\caption{\label{fig:cornerplot} Correlation diagrams of the probability distributions of the best-fit parameters from the MCMC modelling of the ellipsoidal light curve. The contours in the 2D plots show the $68$, $95$ and $99.7$ per cent confidence regions. The right panels show the projected 1D distributions of the parameters, where we mark the mean (solid line) and $68$ per cent confidence level intervals (dashed lines). $M_1$ and $M_2$ are inferred values from $q$, $K_2$, $P$ and $i$.}
\end{figure*}

We used the Markov chain Monte Carlo (MCMC) {\sc emcee}\footnote{\url{https://emcee.readthedocs.io/en/stable/}} package \citep{foreman-13} in {\sc python} along with wide uniform uninformative priors for $\beta$ and $R_{\mathrm{in}}$. The prior for $L_\mathrm{D}$ was flat in log space to allow for an even sampling of the parameter space across orders of magnitude. The absence of eclipses in \targ\ \citep{reinsch-94} implies $i \lesssim 73^{\circ}$ (MR02) and the Chandrasekhar mass limit for a WD imposes $i \gtrsim 55^{\circ}$. We used a flat prior for $\cos i$ and conservatively adopted a $50^{\circ}-85^{\circ}$ range. We adopted Gaussian priors for $K_2$ and $q$, with the mean and standard deviation values obtained in Sections~\ref{sec-k2} and~\ref{sec-vsini}, respectively. We ran the MCMC sampler for 10000 steps with 40 walkers and discarded the first $50$~per cent as burn-in. In each iteration, the comparison between the synthetic and the actual $R$-band light curves is based on the likelihood function of a continuous distribution, computed after normalizing the synthetic light curve to be at the same flux level as the observed one. After several trials we mostly found flat/wide posteriors for $R_{\mathrm{in}}$ and $\beta$, so we were unable to constrain them. Therefore, we decided to marginalise over these two parameters and  provide the correlation plot (Fig.~\ref{fig:cornerplot}) for the remaining parameters and the inferred ones ($M_1$ and $M_2$), whose posterior distributions are close to normal. Table~\ref{tab:parameters_lcm_mcmc} provides the fixed and the fitted model parameters, with the quoted uncertainties established from the $68 $th percentiles in the distributions.

\begin{figure*}
\centering 
\includegraphics[height=8.5cm]{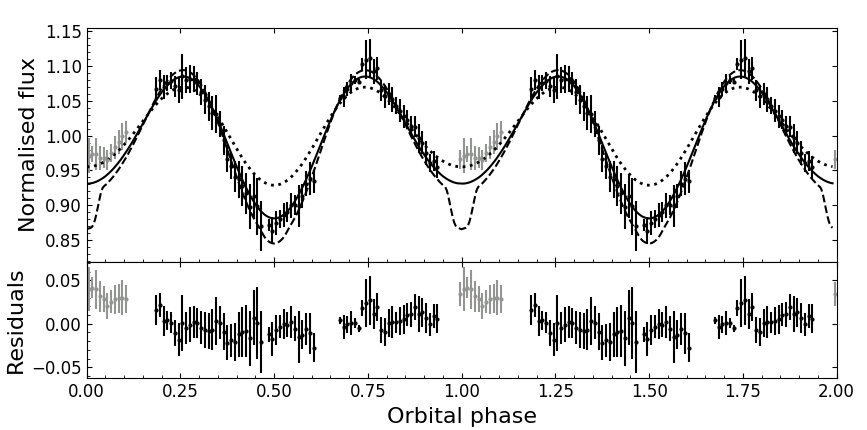}
\caption{\label{fig:lc_modelled} Top panel: Phase-folded $R$-band light curve and the best-fit synthetic model (solid black line). The points in grey colour (phases $0.0-0.1$) have been masked during the fit. The flux of the accretion disc (not shown) in the model is constant along the whole orbit, while the fractional contribution of the donor star to the relative flux is $f=0.73 \pm 0.02$. The dashed and dotted black lines represent the synthetic light curves for $i=73^{\circ}$ and $i=55^{\circ}$, respectively. Bottom panel: Residuals from the fit.}
\end{figure*}

The best-fit model light curve ($\chi^{2}/\mathrm{dof} = 0.6$) is presented in the top panel of Fig.~\ref{fig:lc_modelled} as a solid line. It provided a donor star fractional contribution to the $R$-band flux $f=0.73 \pm 0.02$, in agreement with what we found from the spectroscopy using the K0~III-IV template on the 2017 December 7--10 data (see Section~\ref{sec-vsini}). This template has an effective temperature of $5056 \pm 111 \, \mathrm{K}$, fully consistent with the $T_2$ adopted in the model, and $\mathrm{log} \, g = 3.08 \pm 0.06$~dex \citep{apogee-2020}. The MCMC analysis yields an orbital inclination of $i = 67^{\circ} \pm 5^{\circ}$. In Fig.~\ref{fig:lc_modelled} we show two synthetic light curves computed using the best-fit parameters and the above inclination limits of $55^{\circ}$ and $73^{\circ}$ (dotted and dashed lines, respectively). The $82^{\circ}-86^{\circ}$ binary inclination obtained by \cite{kim-92} from modelling of the dwarf nova outbursts can be rejected, while the estimate of $66^{\circ}$ reported in \cite{bianchini-1982} is in line with our measurement. From our value of the orbital inclination we derive the following binary masses:\par
~\par
$M_{1} = 1.03^{+0.16}_{-0.11} \, \mathrm{M}_{\odot}$~, \, \, \, \, \, \, \, \, \, \, \, \, \,  $M_2 = 0.39^{+0.07}_{-0.06} \, \mathrm{M}_{\odot}$~.\\

The lack of data points around phase $0$ in our $R$-band light curve does not allow us to firmly discard disc eclipses. However, the \textit{TESS} light curve suggests that they are either too shallow to be detected or absent. We have computed $R$-band synthetic light curves and have assumed that the eclipse depth is the same in both the \textit{TESS} and the $R$ bands. Disc eclipses deep enough to be noticeable in the \textit{TESS} light curve would then be produced for $i \gtrsim 71^{\circ}$. Thus, the binary inclination is most likely $i \lesssim 71^{\circ}$, which means that our upper limit on the uncertainty in $i$ is overestimated by $1^{\circ}$. In turn, this leads to a negligible overestimate of 0.01\,M$_\odot$ for the lower limit on $M_1$. Another source of systematic errors in the masses is a potential incorrect determination of the relative contribution of the donor star and the accretion flow to the $R$-band flux. Using the optimal subtraction technique (Section~\ref{sec-vsini}) we determined $f \simeq 0.66 - 0.70$ or $f \simeq 0.73-0.79$ depending on the template, while the best-fit light curve model yields $f=0.73$. Modifying the phase-folded data to simulate light curves with $f=0.65$ and $f=0.8$ results in $i={66^{\circ}}^{+4^{\circ}}_{-5^{\circ}}$ and $i=68^{\circ} \pm 4^{\circ}$, respectively. These changes in the inclination have an impact on the derived masses lower than the reported statistical uncertainties.  Finally, we have also tested the systematic errors associated to the adopted $T_2$. Fixing the temperature of the donor at either $T_2=4700 \, \mathrm{K}$ or $T_2=5300 \, \mathrm{K}$ results in an orbital inclination of $i={68^{\circ}}^{+4^{\circ}}_{-5^{\circ}}$ and $i={66^{\circ}}^{+6^{\circ}}_{-5^{\circ}}$, respectively. Thus, even considering these less probable values for $T_2$, its effect on the dynamical masses is significantly lower than the statistical uncertainties.

\begin{table}
\caption[]{Fixed and fitted parameters of the $R$-band light curve modelling. The type of prior for each parameter was: flat for $\beta$ and $R_{\mathrm{in}}$, flat in log space for $L_D$, flat in cosine space for $i$ and Gaussian for $K_2$ and $q$ (based on the results presented in Sections~\ref{sec-k2} and~\ref{sec-vsini}).}
\label{tab:parameters_lcm_mcmc}
\centering
\begin{tabular}{l l c }
\hline\noalign{\smallskip}
Parameter & Prior & Best-fit value\\
\hline\noalign{\smallskip}
$P$~(d) & Fixed & 1.996872\\
\noalign{\smallskip}
$\alpha~(^{\circ})$ & Fixed & 6\\
\noalign{\smallskip}
$T_2 (\mathrm{K})$ & Fixed & 5000\\
\noalign{\smallskip}
Disc albedo & Fixed & 0.5\\
\noalign{\smallskip}
Donor albedo & Fixed & 0.5\\
\noalign{\smallskip}
$i~(^{\circ})$ & [50, 85]  & $67 \pm 5$ \\
\noalign{\smallskip}
$\log(L_\mathrm{D})~ \mathrm{(\mathrm{erg}\,\mathrm{s}^{-1})}$ & [33, 35.5] &  $33.8 \pm 0.3$ \\
\noalign{\smallskip}
$q$ & $0.38 \pm 0.03$ & $0.38 \pm 0.03$\\
\noalign{\smallskip}
$K_2$ ($\mathrm{km}\,\mathrm{s}^{-1}$) & $126.4 \pm 0.9$ & $126.4 \pm 0.9$\\
\noalign{\smallskip}
$\beta$ & [-3.0, 3.0] & $-$\\
\noalign{\smallskip}
$R_{\mathrm{in}}/a$ & [0.001, 0.2] & $-$\\
\hline\noalign{\smallskip}
\end{tabular}
\end{table}

\section{Discussion}\label{sec-discuss}

\subsection{Binary masses}\label{discussion-masses}

The mass and radius of the Roche-lobe filling donor in \targ\ are $M_2 = 0.39^{+0.07}_{-0.06} \, \mathrm{M}_{\odot}$ and $R_2 = \frac{P}{2 \pi} \frac{v_\mathrm{rot} \sin  i} {\mathrm{sin} \, i} = 2.26 \pm 0.12 \, R_{\odot}$, respectively, which clearly indicates that the donor star is evolved. In fact, we derive $\log \, g = 3.31 \pm 0.09$~dex, which differs from $\log \, g = 4.0$~dex adopted by \cite{harrison-16} to measure a donor effective temperature $T_2=5000 \, \mathrm{K}$ and a metallicity $[\mathrm{Fe}/\mathrm{H}]=-0.3$\,dex from synthetic template spectra. However, our spectral clasification using empirical subgiant templates in the range $\log \, g = 3-3.6$~dex provides a very similar temperature of $\simeq 4900-5150 \, \mathrm{K}$. The above value for $T_2$ is also supported by observations of evolved field stars with  $\log \, g \simeq 3.2-3.4$~dex and metallicities from $-0.2$ to $-0.4$\,dex that have effective temperatures of $4900-5100 \, \mathrm{K}$ \citep{alves-2015}. On the other hand, if the donor star were a stripped giant with all its properties dependent on the mass of its helium core, we could derive its effective temperature from $R_2$. Following the equations given in \cite{webbink-83} and \cite{king-88} we obtained a core mass $0.179 \pm 0.002 \, \mathrm{M}_{\odot}$, a luminosity $2.16 \pm 0.19 \, \mathrm{L_{\odot}}$ and $T_2 = 4660 \pm 26 \, \mathrm{K}$. This lower $T_2$ could be an indication that the donor star in \targ\ is not a stripped giant as initially suggested by \cite{watson-1985}. This would imply that its initial mass was $\lesssim 1.4 \, \mathrm{M}_{\odot}$ \citep{ziolkowski-2020}. 

Our dynamical study of \targ\ has also provided a WD mass  $M_1 = 1.03^{+0.16}_{-0.11} \, \mathrm{M}_{\odot}$. \cite{zorotovic-2011} found an average mass of $0.83 \pm 0.23 \, \mathrm{M}_{\odot}$ from robust measurements in 32 CVs. This value is significantly higher than the mean of $\approx 0.61-0.64 \, \mathrm{M}_{\odot}$ established for isolated WDs \citep{kepler-2016}. The WD mass in \targ\ is no exception –– it is among the ten more massive ones in \citeauthor{zorotovic-2011}'s sample.

\subsection{Comparison with previous WD mass estimates}

\citet[see also \citealp{hachisu-15}]{hachisu-2007} fitted the $1901$ to $1919$ decay light curve of the nova event. They adopted the model described in \cite{hachisupaper1-2006} that uses a decay law for the optical flux, $F_{\lambda} \propto t^{-  \alpha}$, and considers a spherical ejecta whose optical-to-infrared continuum flux is dominated by free-free emission from the optically thin region. They model the different stages of the flux decay using a sequence of steady-state solutions and find that the decline rate depends weakly on the chemical composition and is very sensitive to the WD mass. They provided a WD mass (hereinafter $M_{1}^\mathrm{nova}$) as a function of the hydrogen content of the ejecta, $X$, as $M_{1}^{\mathrm{nova}} (X) \simeq M_{1}^{\mathrm{nova}} (X=0.55) + 0.5(X-0.55)$, valid  for $0.35 \leq X \leq 0.65$ and where $M_{1}^{\mathrm{nova}}(X=0.55)=1.15 \pm 0.05 \, \mathrm{M}_{\odot}$. Note that according to \cite{pottasch-59} the hydrogen content of the nova ejecta in \targ\ is $X \simeq 0.54$. The WD mass obtained from the nova light curve agrees within 1-$\sigma$ of our dynamical measurement. 

More recently, \cite{shara-18} interpolated the grid of nova models from \cite{yaron-2005} to construct functions for the WD mass and the accretion rate depending on the flux amplitude and the mass-loss time of a given nova. Assuming that the latter is equal to the decline time and fitting these functions to the light curve of Nova Persei 1901 they derived  $M_{1}^{\mathrm{nova}}=1.22 \, \mathrm{M}_{\odot}$ with an estimated statistical uncertainty of $\sim \! 0.1 \, \mathrm{M}_{\odot}$. Our dynamical measurement is also in agreement within 1-$ \sigma$ of this value.

Attempts to derive the WD mass from modelling of X-ray data (hereinafter $M_1^\mathrm{X}$) have been reported by different authors. The simplest X-ray models for IPs assume that the accreted matter falls almost radially onto the WD from infinity. Hence, the temperature of the plasma in the post-shock region, $T_\mathrm{s}$, may reflect the depth of the gravitational potential of the WD \citep{aizu-73}:

\begin{equation}
\label{eq:aizu}
kT_\mathrm{s} = \frac{3}{8} \frac{\mathrm{G} M_{1}^\mathrm{X}}{R_{1}} \mu \mathrm{m}_{\mathrm{H}} = 16 \left(\frac{M_{1}^\mathrm{X}}{0.5\, \mathrm{M}_{\odot}}\right)\left(\frac{R_{1}}{10^9 \, \mathrm{cm}}\right)~(\mathrm{keV})\,,
\end{equation} 
\noindent
where $R_{1}$ is the radius of
the WD, $\mu$ is the mean molecular weight ($=0.615$ for solar abundance plasma) and $\mathrm{m_H}$ is the mass of the hydrogen atom. Given that $R_1$ can be expressed as a function of the WD mass using a mass-radius relation \citep[e.g.][]{nauenberg-72}, the temperature can be expressed as a function of the WD mass only. The hot post-shock region cools mainly via thermal bremsstrahlung emission in the hard X-ray regime. Hence, the easiest way to estimate the plasma temperature and thus derive the WD mass is to fit the hard X-ray spectra with single-temperature bremsstrahlung models and adopt the best-fit temperature as the maximum shock temperature. 

\cite{landi-2009} fitted the combined $0.2-10$~keV \textit{Swift}/XRT and $20-100$~keV \textit{INTEGRAL}/IBIS spectrum of \targ\ with a blackbody plus bremsstrahlung model for the soft and hard X-ray emission, respectively, taking into account the local and interstellar absorption. The best-fit bremsstrahlung temperature $kT_\mathrm{s}=23^{+9.2}_{-6.5} \, \mathrm{keV}$ yields (Equation~\ref{eq:aizu}) $M_1^\mathrm{X} = 0.87^{+0.25}_{-0.21} \, \mathrm{M}_{\odot}$ \citep{brunchsweiger-2009}. Given its large uncertainty this mass estimate agrees within 1-$\sigma$ of our dynamical measurement. 

Other authors have considered a multi-temperature continuum post-shock region. \cite{suleimanov-2005} presented a model with temperature and density as functions of the distance to the WD surface under the assumption that matter is accreted from infinity. The latter is a good approximation when the Alfvén radius ($R_\mathrm{A}$) is more than $10$ times larger than the WD radius. They fitted this model to a $3-100$~keV outburst spectrum of \targ\ obtained by combining \textit{RXTE} PCA and \textit{HEXTE} data and found a much lower WD mass of $M_1^\mathrm{X} = 0.59 \pm 0.05 \, \mathrm{M}_{\odot}$. Given that $R_\mathrm{A}/R_{1} < 5$ when \targ\ is in outburst and $R_\mathrm{A}/R_{1} < 10$ during quiescence, \cite{suleimanov-2005} concluded that their value of $M_1^\mathrm{X}$ obtained from outburst data underestimated the WD mass by at least $20$ per cent. Our dynamical mass measurement allows to establish that their $M_1^\mathrm{X}$ actually underestimates the WD mass by $73^{+40}_{-17}$ per cent. These authors also fitted their models to the $3-20$~keV PCA data alone and obtained $M_1^\mathrm{X} = 1.24 \pm 0.10 \, \mathrm{M}_{\odot}$. They showed that the use of the X-ray continuum in the energy range $< 20$~keV does not provide an accurate estimate of $M_1$ because for WD masses $>0.6 \, \mathrm{M}_{\odot}$ the post-shock temperature is $>20$~keV. Further, the IP X-ray continuum in that range is affected by interstellar and intrinsic absorption and possibly a reflection component. In this regard, \cite{evans-2007} fitted the multi-temperature continuum model for the post-shock region by \cite{cropper-99}, which is similar to that in \cite{suleimanov-2005}, to a $0.2-12$~keV \textit{XMM-Newton}/EPIC-pn spectrum of \targ\ in outburst. The fit provided $M_1^\mathrm{X} = 0.92^{+0.39}_{-0.13} \, \mathrm{M}_{\odot}$, which agrees with our measurement given the large uncertainties.

The systematic underestimate of the WD mass of \targ\ as a function of its accretion state was explored by \cite{brunchsweiger-2009}. They fitted the \citeauthor{suleimanov-2005} model to both quiescence and outburst $15-200$~keV \textit{Swift}/BAT spectra. For the outburst data they derived a mass $M_1^\mathrm{X}=0.67-0.74 \, \mathrm{M}_{\odot}$ lower than the result in quiescence $M_1^X = 0.90 \pm 0.12 \, \mathrm{M}_{\odot}$. They suggested a true WD mass of $1.15 \, \mathrm{M}_{\odot}$ from fitting the relation between $M_1^X$ and the accretion rate of the system when the data are taken. This value is in line with our result.

In order to determine a $M_1^\mathrm{X}$ free of this underestimate effect, \cite{suleimanov-16} proposed a method that obtains $M_1^\mathrm{X}$ and $R_\mathrm{A}$ simultaneously from the break frequency of the power spectrum\footnote{The power spectra of X-ray pulsars and IPs exhibit a break at the frequency of the Keplerian rotation at the boundary of the magnetosphere \citep{revnivtsev-2009, revnivtsev-2011}.} and a model with the matter falling from $R_\mathrm{A}$, not from infinity. By fitting this refined model to a $20-80$ keV \textit{NuSTAR} spectrum of \targ\ in outburst they found $M_1^\mathrm{X} = 0.86 \pm 0.02 \, \mathrm{M}_{\odot}$ and $R_\textit{A}/R_{1} = 2.8 \pm 0.2$. This same method was applied to a $20-80$ keV \textit{NuSTAR} spectrum taken in quiescence \citep{suleimanov-19} that provided $M_1^\mathrm{X} = 0.79 \pm 0.01 \, \mathrm{M}_{\odot}$ and $R_\mathrm{A}/R_{1} = 3.18 \pm 0.17$. These $M_1^\mathrm{X}$ values are consistent within 2$-\sigma$ of the dynamical WD mass presented in this work.\\

Other methods have been explored to derive the WD mass in \targ\ using X-ray data. \cite{ezuka-99} fitted the continuum emission of a $0.5-10$~keV spectrum of \targ\ in outburst combining \textit{ASCA}/SIS and GIS data. Their model consists of  single-temperature thermal bremsstrahlung that undergoes
multi-column absorption. This included the fluorescent ($6.4$~keV) and plasma ($6.7$~and $7.0$~keV) components of the iron K$\alpha$ emission line. They used the intensity ratios of these lines to measure the ionisation temperature and obtained $kT = 7.8^{+2.1}_{-1.8}$~keV. They also derived a relation between this and the corrected temperature of the continuum (see panel $b$ of fig.~$4$ in \citealt{ezuka-99}). This yields $M_1^\mathrm{X} = 0.52^{+0.34}_{-0.16} \, \mathrm{M}_{\odot}$ ($90$ per cent confidence level). The discrepancy with our $1.03^{+0.16}_{-0.11}\,\mathrm{M}_\odot$ dynamical value shows that the above temperature relation must be improved in order to obtain reliable WD masses. Finally, \cite{wada-2018} derived a $M_{1}^\mathrm{X}$--$R_\mathrm{A}$ relation for a given temperature of the post-shock region by modifying Equation~\ref{eq:aizu} to account for the free-fall initial point and the shock height. They fitted a multi-temperature optically-thin plasma model to the quiescence and outburst \textit{NuSTAR} spectra analysed in \cite{suleimanov-16,suleimanov-19}. From the flux and plasma temperature measured at different accretion states they established that $R_\mathrm{A}$ in quiescence is $3.9$ times larger than during outburst. By finding the $M_1^\mathrm{X}$ value that satisfies this condition they derived $0.87 \pm 0.08 \, \mathrm{M}_{\odot}$, which is consistent within 1-$\sigma$ of our dynamical mass.

\section{Conclusions}\label{sec-conclusions}

\begin{table}
\caption[]{Fundamental parameters of \targ\ obtained in this work.}
\label{tab:parameters_summary}
\centering
\begin{tabular}{l c }
\hline\noalign{\smallskip}
Parameter & Value\\
\hline\noalign{\smallskip}
$P \, (\mathrm{d})$ & $1.996872 \pm 0.000009$ \\
\noalign{\smallskip}
$K_2 \, (\mathrm{km~s^{-1}})$ & $126.4 \pm 0.9$ \\
\noalign{\smallskip}
$v_\mathrm{rot} \sin  i \, (\mathrm{km~s^{-1}})$ & $52 \pm 2$ \\
\noalign{\smallskip}
$q$ & $0.38 \pm 0.03$ \\
\noalign{\smallskip}
$i \,(^{\circ})$ & $67 \pm 5$ \\
\noalign{\smallskip}
$M_1 \, (\mathrm{M}_{\odot})$ & $1.03^{+0.16}_{-0.11}$\\
\noalign{\smallskip}
$M_2 \, (\mathrm{M}_{\odot})$ & $0.39^{+0.07}_{-0.06}$\\
\noalign{\smallskip}
$R_2 \, (\mathrm{R}_{\odot})$ & $2.26 \pm 0.11$\\
\hline\noalign{\smallskip}
\end{tabular}
\end{table}

We have presented a dynamical study of the intermediate polar cataclysmic variable \targ\ 120 years after its nova explosion. We obtained the radial velocity curve of the donor star and refined the orbital period of the system to $1.996872 \pm 0.000009$~d and $K_2$ to $126.4 \pm 0.9 \, \mathrm{km~s^{-1}}$. From our higher resolution spectroscopy we established a rotational broadening for the donor star absorption lines $v_\mathrm{rot} \sin  i = 52 \pm 2 \, \mathrm{km~s^{-1}}$. Partially simultaneous photometry and spectroscopy allowed us to construct a phase-folded $R$-band light curve as free as possible from night-to-night accretion variability. Modelling of the light curve allowed to establish the binary orbital inclination $i=67 \pm 5^{\circ}$. We derived the dynamical masses $1.03^{+0.16}_{-0.11} \, \mathrm{M}_{\odot}$ and $0.39^{+0.07}_{-0.06} \, \mathrm{M}_{\odot}$ for the WD and the evolved donor star, respectively.

The WD in \targ\ has one of the highest masses confirmed dynamically for a CV and our value is in agreement with those obtained from modelling of the nova decay light curve. We also compared the WD mass with X-ray modelling estimates. We confirm that the WD mass values derived from the hard X-ray spectra ($> 20$~keV) continuum temperature assuming matter free falling from infinity are significantly underestimated. Models considering gas free falling from a finite distance \citep{suleimanov-16, suleimanov-19} agree with our measurement within 2-$\sigma$. The same degree of agreement is found for the estimates obtained from the intensity ratio of the fluorescent and plasma components of the K$\alpha$ emission line \citep{ezuka-99}. On the other hand, the WD mass value obtained by \cite{wada-2018} using the quiescence-to-outburst Alfvén radius ratio is consistent within 1-$\sigma$ of our dynamical measurement.

Our robust dynamical mass for the WD in \targ\ has served as a stringent test for the values reported by other authors using indirect methods. However, a more accurate $M_1$ is still needed to further constrain the nova decay light curve and X-ray spectral modelling techniques. Future multi-colour photometry sampling the full binary orbit will likely result in a more precise value of the inclination that could potentially lead to a refined dynamical WD mass. 

Finally, our study supports a subgiant donor with an effective temperature of $\simeq 5000$~K, in line with the values obtained from infrared spectroscopy by \cite{harrison-2013} and \cite{harrison-16}. The stripped giant model failed to explain the observed temperature, which suggests that the initial mass of the donor star was $\lesssim 1.4 \, \mathrm{M}_{\odot}$.

\section*{Acknowledgements} 

We thank the anonymous referee for a careful review of the manuscript and useful comments. MAPT and PR-G acknowledge support from the State Research Agency (AEI) of the Spanish Ministry of Science, Innovation and Universities (MCIU) and the European Regional Development Fund (FEDER) under grants AYA2017-83216-P and AYA2017--83383--P, respectively. MAPT was also supported by a Ram\'on y Cajal Fellowship RYC-2015-17854. AR acknowledges the Research Associate Fellowship with order no. 03(1428)/18/EMR-II under Council of Scientific and Industrial Research (CSIR). JJR is thankful for support from NSFC (grant No. 11903048, 11833006).  IPM acknowledges funding from the Netherlands Research School for Astronomy (grant no. NOVA5-NW3-10.3.5.14). Based on observations made with the Nordic Optical Telescope, owned in collaboration by the University of Turku and Aarhus University, and operated jointly by Aarhus University, the University of Turku and the University of Oslo, representing Denmark, Finland and Norway, the University of Iceland and Stockholm University at the Observatorio del Roque de los Muchachos, La Palma, Spain, of the Instituto de Astrof\'\i sica de Canarias. This  article  is  also  based  on  observations  made with  the  WHT, operated on the island of La Palma by the Isaac Newton Group of Telescopes in the Spanish Observatorio del Roque de los Muchachos of the Instituto de Astrof\'\i sica de Canarias. We thank the staff of VBO, Kavalur, IAO, Hanle and CREST, Hosakote that made the 1.3-m JCBT and the 2-m HCT observations possible. The facilities at VBO, IAO and CREST are operated by the Indian Institute of Astrophysics, Bengaluru. We acknowledge the support of the staff of the Xinglong 2.16-m telescope. This work was partially supported by the Open Project Program of the Key Laboratory of Optical Astronomy, National Astronomical Observatories, Chinese Academy of Sciences. This paper includes data collected by the TESS mission. Funding for the TESS mission is provided by the NASA's Science Mission Directorate. \textsc{iraf} is distributed by the National OpticalAstronomy Observatory, which is operated by the Association of Universities for Research in Astronomy (AURA) under a cooperative agreement with the National Science Foundation, USA. This research has made use of the APASS database, located at the AAVSO website.  Funding for APASS has been provided by the Robert Martin Ayers Sciences Fund. We also have made use of the  ‘Aladin Sky Atlas’ developed at CDS, Strasbourg Observatory, France. The use of the \textsc{pamela} and \textsc{molly} packages developed by Tom Marsh is acknowledged.

\section*{Data availability} 

The NOT and WHT data are available at \url{http://www.not.iac.es/observing/forms/fitsarchive/} and \url{http://casu.ast.cam.ac.uk/casuadc/ingarch/query}, respectively. They can be searched by observing date and object coordinates. The 0.3-m SC30, 0.43-m CDK, 0.4-m UOAO and 2.16-m Xinglong data are available from the corresponding author on request. The HCT and 1.3-m JCBT data will be made available by G. C. Anupama and M. Pavana on request.



\bibliographystyle{mnras} 
\bibliography{bibliography} 

\begin{thebibliography}{}
\makeatletter
\relax
\def\mn@urlcharsother{\let\do\@makeother \do\$\do\&\do\#\do\^\do\_\do\%\do\~}
\def\mn@doi{\begingroup\mn@urlcharsother \@ifnextchar [ {\mn@doi@}
  {\mn@doi@[]}}
\def\mn@doi@[#1]#2{\def\@tempa{#1}\ifx\@tempa\@empty \href
  {http://dx.doi.org/#2} {doi:#2}\else \href {http://dx.doi.org/#2} {#1}\fi
  \endgroup}
\def\mn@eprint#1#2{\mn@eprint@#1:#2::\@nil}
\def\mn@eprint@arXiv#1{\href {http://arxiv.org/abs/#1} {{\tt arXiv:#1}}}
\def\mn@eprint@dblp#1{\href {http://dblp.uni-trier.de/rec/bibtex/#1.xml}
  {dblp:#1}}
\def\mn@eprint@#1:#2:#3:#4\@nil{\def\@tempa {#1}\def\@tempb {#2}\def\@tempc
  {#3}\ifx \@tempc \@empty \let \@tempc \@tempb \let \@tempb \@tempa \fi \ifx
  \@tempb \@empty \def\@tempb {arXiv}\fi \@ifundefined
  {mn@eprint@\@tempb}{\@tempb:\@tempc}{\expandafter \expandafter \csname
  mn@eprint@\@tempb\endcsname \expandafter{\@tempc}}}

\bibitem[\protect\citeauthoryear{{Aizu}}{{Aizu}}{1973}]{aizu-73}
{Aizu} K.,  1973, \mn@doi [Progress of Theoretical Physics]
  {10.1143/PTP.49.1184}, \href
  {https://ui.adsabs.harvard.edu/abs/1973PThPh..49.1184A} {49, 1184}

\bibitem[\protect\citeauthoryear{{Alves} et~al.,}{{Alves}
  et~al.}{2015}]{alves-2015}
{Alves} S.,  et~al., 2015, \mn@doi [\mnras] {10.1093/mnras/stv189}, \href
  {https://ui.adsabs.harvard.edu/abs/2015MNRAS.448.2749A} {448, 2749}

\bibitem[\protect\citeauthoryear{{Anupama} \& {Kantharia}}{{Anupama} \&
  {Kantharia}}{2005}]{anupama-2005}
{Anupama} G.~C.,  {Kantharia} N.~G.,  2005, \mn@doi [\aap]
  {10.1051/0004-6361:20042371}, \href
  {https://ui.adsabs.harvard.edu/abs/2005A&A...435..167A} {435, 167}

\bibitem[\protect\citeauthoryear{{Anupama} \& {Prabhu}}{{Anupama} \&
  {Prabhu}}{1993}]{anupama-93}
{Anupama} G.~C.,  {Prabhu} T.~P.,  1993, \mn@doi [\mnras]
  {10.1093/mnras/263.2.335}, \href
  {https://ui.adsabs.harvard.edu/abs/1993MNRAS.263..335A} {263, 335}

\bibitem[\protect\citeauthoryear{{Bayless}, {Robinson}, {Hynes}, {Ashcraft}  \&
  {Cornell}}{{Bayless} et~al.}{2010}]{bayless-2010}
{Bayless} A.~J.,  {Robinson} E.~L.,  {Hynes} R.~I.,  {Ashcraft} T.~A.,
  {Cornell} M.~E.,  2010, \mn@doi [\apj] {10.1088/0004-637X/709/1/251}, \href
  {https://ui.adsabs.harvard.edu/abs/2010ApJ...709..251B} {709, 251}

\bibitem[\protect\citeauthoryear{{Bianchini} \& {Sabbadin}}{{Bianchini} \&
  {Sabbadin}}{1983}]{bianchini-83}
{Bianchini} A.,  {Sabbadin} F.,  1983, {The continuum energy distribution of
  the old-nova GK Per (1901)}.
pp 127--130, \mn@doi{10.1007/978-94-009-7118-9\_15}

\bibitem[\protect\citeauthoryear{{Bianchini}, {Sabbadin}  \&
  {Hamzaoglu}}{{Bianchini} et~al.}{1982}]{bianchini-1982}
{Bianchini} A.,  {Sabbadin} F.,   {Hamzaoglu} E.,  1982, \aap, \href
  {https://ui.adsabs.harvard.edu/abs/1982A&A...106..176B} {106, 176}

\bibitem[\protect\citeauthoryear{{Bode}, {Seaquist}, {Frail}, {Roberts},
  {Whittet}, {Evans}  \& {Albinson}}{{Bode} et~al.}{1987}]{bode-87}
{Bode} M.~F.,  {Seaquist} E.~R.,  {Frail} D.~A.,  {Roberts} J.~A.,  {Whittet}
  D.~C.~B.,  {Evans} A.,   {Albinson} J.~S.,  1987, \mn@doi [\nat]
  {10.1038/329519a0}, \href
  {https://ui.adsabs.harvard.edu/abs/1987Natur.329..519B} {329, 519}

\bibitem[\protect\citeauthoryear{{Brasseur}, {Phillip}, {Fleming}, {Mullally}
  \& {White}}{{Brasseur} et~al.}{2019}]{brasseur-19}
{Brasseur} C.~E.,  {Phillip} C.,  {Fleming} S.~W.,  {Mullally} S.~E.,   {White}
  R.~L.,  2019, {Astrocut: Tools for creating cutouts of TESS images}
  (\mn@eprint {ascl} {1905.007})

\bibitem[\protect\citeauthoryear{{Brewer}, {Fischer}, {Valenti}  \&
  {Piskunov}}{{Brewer} et~al.}{2016}]{brewer-2016}
{Brewer} J.~M.,  {Fischer} D.~A.,  {Valenti} J.~A.,   {Piskunov} N.,  2016,
  \mn@doi [\apjs] {10.3847/0067-0049/225/2/32}, \href
  {https://ui.adsabs.harvard.edu/abs/2016ApJS..225...32B} {225, 32}

\bibitem[\protect\citeauthoryear{{Brunschweiger}, {Greiner}, {Ajello}  \&
  {Osborne}}{{Brunschweiger} et~al.}{2009}]{brunchsweiger-2009}
{Brunschweiger} J.,  {Greiner} J.,  {Ajello} M.,   {Osborne} J.,  2009, \mn@doi
  [\aap] {10.1051/0004-6361/200811285}, \href
  {https://ui.adsabs.harvard.edu/abs/2009A&A...496..121B} {496, 121}

\bibitem[\protect\citeauthoryear{{Chambers} et~al.,}{{Chambers}
  et~al.}{2016}]{chambers-2016}
{Chambers} K.~C.,  et~al., 2016, arXiv e-prints, \href
  {https://ui.adsabs.harvard.edu/abs/2016arXiv161205560C} {p. arXiv:1612.05560}

\bibitem[\protect\citeauthoryear{{Chanmugam} \& {Wagner}}{{Chanmugam} \&
  {Wagner}}{1977}]{chanmugam-77}
{Chanmugam} G.,  {Wagner} R.~L.,  1977, \mn@doi [\apjl] {10.1086/182399}, \href
  {https://ui.adsabs.harvard.edu/abs/1977ApJ...213L..13C} {213, L13}

\bibitem[\protect\citeauthoryear{{Claret}}{{Claret}}{2000a}]{claret-2000-gd}
{Claret} A.,  2000a, \aap, \href
  {https://ui.adsabs.harvard.edu/abs/2000A&A...359..289C} {359, 289}

\bibitem[\protect\citeauthoryear{{Claret}}{{Claret}}{2000b}]{claret-2000-ld}
{Claret} A.,  2000b, \aap, \href
  {https://ui.adsabs.harvard.edu/abs/2000A&A...363.1081C} {363, 1081}

\bibitem[\protect\citeauthoryear{{Claret}, {Diaz-Cordoves}  \&
  {Gimenez}}{{Claret} et~al.}{1995}]{claret95}
{Claret} A.,  {Diaz-Cordoves} J.,   {Gimenez} A.,  1995, \aaps, \href
  {https://ui.adsabs.harvard.edu/abs/1995A&AS..114..247C} {114, 247}

\bibitem[\protect\citeauthoryear{{Cox}}{{Cox}}{2000}]{cox-2000}
{Cox} A.~N.,  2000, {Allen's astrophysical quantities}.
Springer-Verlag, New York

\bibitem[\protect\citeauthoryear{{Crampton}, {Cowley}  \& {Fisher}}{{Crampton}
  et~al.}{1986}]{crampton-1986}
{Crampton} D.,  {Cowley} A.~P.,   {Fisher} W.~A.,  1986, \mn@doi [\apj]
  {10.1086/163856}, \href
  {https://ui.adsabs.harvard.edu/abs/1986ApJ...300..788C} {300, 788}

\bibitem[\protect\citeauthoryear{{Cropper}}{{Cropper}}{1990}]{cropper-90}
{Cropper} M.,  1990, \mn@doi [\ssr] {10.1007/BF00177799}, \href
  {https://ui.adsabs.harvard.edu/abs/1990SSRv...54..195C} {54, 195}

\bibitem[\protect\citeauthoryear{{Cropper}, {Wu}, {Ramsay}  \&
  {Kocabiyik}}{{Cropper} et~al.}{1999}]{cropper-99}
{Cropper} M.,  {Wu} K.,  {Ramsay} G.,   {Kocabiyik} A.,  1999, \mn@doi [\mnras]
  {10.1046/j.1365-8711.1999.02570.x}, \href
  {https://ui.adsabs.harvard.edu/abs/1999MNRAS.306..684C} {306, 684}

\bibitem[\protect\citeauthoryear{{Dougherty}, {Waters}, {Bode}, {Lloyd},
  {Kester}  \& {Bontekoe}}{{Dougherty} et~al.}{1996}]{dougherty-96}
{Dougherty} S.~M.,  {Waters} L.~B.~F.~M.,  {Bode} M.~F.,  {Lloyd} H.~M.,
  {Kester} D.~J.~M.,   {Bontekoe} T.~R.,  1996, \aap, \href
  {https://ui.adsabs.harvard.edu/abs/1996A&A...306..547D} {306, 547}

\bibitem[\protect\citeauthoryear{{Eggleton}}{{Eggleton}}{1983}]{eggleton-83}
{Eggleton} P.~P.,  1983, \mn@doi [\apj] {10.1086/160960}, \href
  {https://ui.adsabs.harvard.edu/abs/1983ApJ...268..368E} {268, 368}

\bibitem[\protect\citeauthoryear{{Evans} \& {Hellier}}{{Evans} \&
  {Hellier}}{2007}]{evans-2007}
{Evans} P.~A.,  {Hellier} C.,  2007, \mn@doi [\apj] {10.1086/518552}, \href
  {https://ui.adsabs.harvard.edu/abs/2007ApJ...663.1277E} {663, 1277}

\bibitem[\protect\citeauthoryear{{Ezuka} \& {Ishida}}{{Ezuka} \&
  {Ishida}}{1999}]{ezuka-99}
{Ezuka} H.,  {Ishida} M.,  1999, \mn@doi [\apjs] {10.1086/313181}, \href
  {https://ui.adsabs.harvard.edu/abs/1999ApJS..120..277E} {120, 277}

\bibitem[\protect\citeauthoryear{{Faulkner}, {Flannery}  \&
  {Warner}}{{Faulkner} et~al.}{1972}]{faulkneretal72}
{Faulkner} J.,  {Flannery} B.~P.,   {Warner} B.,  1972, \mn@doi [\apjl]
  {10.1086/180989}, \href
  {https://ui.adsabs.harvard.edu/abs/1972ApJ...175L..79F} {175, L79}

\bibitem[\protect\citeauthoryear{{Foreman-Mackey}, {Hogg}, {Lang}  \&
  {Goodman}}{{Foreman-Mackey} et~al.}{2013}]{foreman-13}
{Foreman-Mackey} D.,  {Hogg} D.~W.,  {Lang} D.,   {Goodman} J.,  2013, \mn@doi
  [\pasp] {10.1086/670067}, \href
  {http://adsabs.harvard.edu/abs/2013PASP..125..306F} {125, 306}

\bibitem[\protect\citeauthoryear{{Gaia Collaboration} et~al.,}{{Gaia
  Collaboration} et~al.}{2018}]{gaia}
{Gaia Collaboration} et~al., 2018, \mn@doi [Astronomy \& Astrophysics]
  {10.1051/0004-6361/201833051}, \href
  {https://ui.adsabs.harvard.edu/abs/2018A&A...616A...1G} {616, A1}

\bibitem[\protect\citeauthoryear{{Gallagher} \& {Oinas}}{{Gallagher} \&
  {Oinas}}{1974}]{gallagher-74}
{Gallagher} J.~S.,  {Oinas} V.,  1974, \mn@doi [\pasp] {10.1086/129704}, \href
  {https://ui.adsabs.harvard.edu/abs/1974PASP...86..952G} {86, 952}

\bibitem[\protect\citeauthoryear{{Garlick}, {Mittaz}, {Rosen}  \&
  {Mason}}{{Garlick} et~al.}{1994}]{garlick-94}
{Garlick} M.~A.,  {Mittaz} J.~P.~D.,  {Rosen} S.~R.,   {Mason} K.~O.,  1994,
  \mn@doi [\mnras] {10.1093/mnras/269.2.517}, \href
  {https://ui.adsabs.harvard.edu/abs/1994MNRAS.269..517G} {269, 517}

\bibitem[\protect\citeauthoryear{{Gazeas}}{{Gazeas}}{2016}]{kosmas-2016}
{Gazeas} K.,  2016, in Revista Mexicana de Astronomia y Astrofisica Conference
  Series. pp 22--23

\bibitem[\protect\citeauthoryear{{Gray}}{{Gray}}{1992}]{libro-gray}
{Gray} D.~F.,  1992, {The Observation and Analysis of Stellar Photospheres}.
Cambridge University Press, Cambridge

\bibitem[\protect\citeauthoryear{{Hachisu} \& {Kato}}{{Hachisu} \&
  {Kato}}{2006}]{hachisupaper1-2006}
{Hachisu} I.,  {Kato} M.,  2006, \mn@doi [\apjs] {10.1086/508063}, \href
  {https://ui.adsabs.harvard.edu/abs/2006ApJS..167...59H} {167, 59}

\bibitem[\protect\citeauthoryear{{Hachisu} \& {Kato}}{{Hachisu} \&
  {Kato}}{2007}]{hachisu-2007}
{Hachisu} I.,  {Kato} M.,  2007, \mn@doi [\apj] {10.1086/517600}, \href
  {https://ui.adsabs.harvard.edu/abs/2007ApJ...662..552H} {662, 552}

\bibitem[\protect\citeauthoryear{{Hachisu} \& {Kato}}{{Hachisu} \&
  {Kato}}{2015}]{hachisu-15}
{Hachisu} I.,  {Kato} M.,  2015, \mn@doi [\apj] {10.1088/0004-637X/798/2/76},
  \href {https://ui.adsabs.harvard.edu/abs/2015ApJ...798...76H} {798, 76}

\bibitem[\protect\citeauthoryear{{Harrison}}{{Harrison}}{2016}]{harrison-16}
{Harrison} T.~E.,  2016, \mn@doi [\apj] {10.3847/0004-637X/833/1/14}, \href
  {https://ui.adsabs.harvard.edu/abs/2016ApJ...833...14H} {833, 14}

\bibitem[\protect\citeauthoryear{{Harrison} \& {Hamilton}}{{Harrison} \&
  {Hamilton}}{2015}]{harrison-2015}
{Harrison} T.~E.,  {Hamilton} R.~T.,  2015, \mn@doi [\aj]
  {10.1088/0004-6256/150/5/142}, \href
  {https://ui.adsabs.harvard.edu/abs/2015AJ....150..142H} {150, 142}

\bibitem[\protect\citeauthoryear{{Harrison}, {Bornak}, {McArthur}  \&
  {Benedict}}{{Harrison} et~al.}{2013}]{harrison-2013}
{Harrison} T.~E.,  {Bornak} J.,  {McArthur} B.~E.,   {Benedict} G.~F.,  2013,
  \mn@doi [\apj] {10.1088/0004-637X/767/1/7}, \href
  {https://ui.adsabs.harvard.edu/abs/2013ApJ...767....7H} {767, 7}

\bibitem[\protect\citeauthoryear{{Henden} \& {Honeycutt}}{{Henden} \&
  {Honeycutt}}{1995}]{hh-95}
{Henden} A.~A.,  {Honeycutt} R.~K.,  1995, \mn@doi [\pasp] {10.1086/133557},
  \href {https://ui.adsabs.harvard.edu/abs/1995PASP..107..324H} {107, 324}

\bibitem[\protect\citeauthoryear{{Henden} \& {Honeycutt}}{{Henden} \&
  {Honeycutt}}{1997}]{hh-97}
{Henden} A.~A.,  {Honeycutt} R.~K.,  1997, \mn@doi [\pasp] {10.1086/133901},
  \href {https://ui.adsabs.harvard.edu/abs/1997PASP..109..441H} {109, 441}

\bibitem[\protect\citeauthoryear{{Hudec}}{{Hudec}}{1981}]{hudec-1981}
{Hudec} R.,  1981, Bulletin of the Astronomical Institutes of Czechoslovakia,
  \href {https://ui.adsabs.harvard.edu/abs/1981BAICz..32...93H} {32, 93}

\bibitem[\protect\citeauthoryear{{J{\"o}nsson} et~al.,}{{J{\"o}nsson}
  et~al.}{2020}]{apogee-2020}
{J{\"o}nsson} H.,  et~al., 2020, \mn@doi [\aj] {10.3847/1538-3881/aba592},
  \href {https://ui.adsabs.harvard.edu/abs/2020AJ....160..120J} {160, 120}

\bibitem[\protect\citeauthoryear{{Kepler} et~al.,}{{Kepler}
  et~al.}{2016}]{kepler-2016}
{Kepler} S.~O.,  et~al., 2016, \mn@doi [\mnras] {10.1093/mnras/stv2526}, \href
  {https://ui.adsabs.harvard.edu/abs/2016MNRAS.455.3413K} {455, 3413}

\bibitem[\protect\citeauthoryear{{Kim}, {Wheeler}  \& {Mineshige}}{{Kim}
  et~al.}{1992}]{kim-92}
{Kim} S.-W.,  {Wheeler} J.~C.,   {Mineshige} S.,  1992, \mn@doi [\apj]
  {10.1086/170870}, \href
  {https://ui.adsabs.harvard.edu/abs/1992ApJ...384..269K} {384, 269}

\bibitem[\protect\citeauthoryear{{King}}{{King}}{1988}]{king-88}
{King} A.~R.,  1988, \qjras, \href
  {https://ui.adsabs.harvard.edu/abs/1988QJRAS..29....1K} {29, 1}

\bibitem[\protect\citeauthoryear{{Koen}, {Kilkenny}, {van Wyk}  \&
  {Marang}}{{Koen} et~al.}{2010}]{koen-2010}
{Koen} C.,  {Kilkenny} D.,  {van Wyk} F.,   {Marang} F.,  2010, \mn@doi
  [\mnras] {10.1111/j.1365-2966.2009.16182.x}, \href
  {https://ui.adsabs.harvard.edu/abs/2010MNRAS.403.1949K} {403, 1949}

\bibitem[\protect\citeauthoryear{{Kraft}}{{Kraft}}{1964}]{kraft-64}
{Kraft} R.~P.,  1964, \mn@doi [\apj] {10.1086/147776}, \href
  {https://ui.adsabs.harvard.edu/abs/1964ApJ...139..457K} {139, 457}

\bibitem[\protect\citeauthoryear{{Kurucz}}{{Kurucz}}{1996}]{kurucz-96}
{Kurucz} R.~L.,  1996, in {Adelman} S.~J.,  {Kupka} F.,   {Weiss} W.~W.,  eds,
  Astronomical Society of the Pacific Conference Series Vol. 108, M.A.S.S.,
  Model Atmospheres and Spectrum Synthesis. p.~2

\bibitem[\protect\citeauthoryear{{Landi}, {Bassani}, {Dean}, {Bird}, {Fiocchi},
  {Bazzano}, {Nousek}  \& {Osborne}}{{Landi} et~al.}{2009}]{landi-2009}
{Landi} R.,  {Bassani} L.,  {Dean} A.~J.,  {Bird} A.~J.,  {Fiocchi} M.,
  {Bazzano} A.,  {Nousek} J.~A.,   {Osborne} J.~P.,  2009, \mn@doi [\mnras]
  {10.1111/j.1365-2966.2008.14086.x}, \href
  {https://ui.adsabs.harvard.edu/abs/2009MNRAS.392..630L} {392, 630}

\bibitem[\protect\citeauthoryear{{Lightkurve Collaboration}
  et~al.,}{{Lightkurve Collaboration} et~al.}{2018}]{lightkurve-18}
{Lightkurve Collaboration} et~al., 2018, {Lightkurve: Kepler and TESS time
  series analysis in Python} (\mn@eprint {ascl} {1812.013})

\bibitem[\protect\citeauthoryear{{Liimets}, {Corradi}, {Santander-Garc{\'\i}a},
  {Villaver}, {Rodr{\'\i}guez-Gil}, {Verro}  \& {Kolka}}{{Liimets}
  et~al.}{2012}]{liimets-2012}
{Liimets} T.,  {Corradi} R.~L.~M.,  {Santander-Garc{\'\i}a} M.,  {Villaver} E.,
   {Rodr{\'\i}guez-Gil} P.,  {Verro} K.,   {Kolka} I.,  2012, \mn@doi [\apj]
  {10.1088/0004-637X/761/1/34}, \href
  {https://ui.adsabs.harvard.edu/abs/2012ApJ...761...34L} {761, 34}

\bibitem[\protect\citeauthoryear{{Luck}}{{Luck}}{2015}]{luck-2015}
{Luck} R.~E.,  2015, \mn@doi [\aj] {10.1088/0004-6256/150/3/88}, \href
  {https://ui.adsabs.harvard.edu/abs/2015AJ....150...88L} {150, 88}

\bibitem[\protect\citeauthoryear{{Marsh}}{{Marsh}}{1989}]{marsh-89}
{Marsh} T.~R.,  1989, \mn@doi [\pasp] {10.1086/132570}, \href
  {https://ui.adsabs.harvard.edu/abs/1989PASP..101.1032M} {101, 1032}

\bibitem[\protect\citeauthoryear{{Marsh}, {Robinson}  \& {Wood}}{{Marsh}
  et~al.}{1994}]{marsh94}
{Marsh} T.~R.,  {Robinson} E.~L.,   {Wood} J.~H.,  1994, \mn@doi [\mnras]
  {10.1093/mnras/266.1.137}, \href
  {https://ui.adsabs.harvard.edu/abs/1994MNRAS.266..137M} {266, 137}

\bibitem[\protect\citeauthoryear{{Morales-Rueda}, {Still}, {Roche}, {Wood}  \&
  {Lockley}}{{Morales-Rueda} et~al.}{2002}]{moralesrueda}
{Morales-Rueda} L.,  {Still} M.~D.,  {Roche} P.,  {Wood} J.~H.,   {Lockley}
  J.~J.,  2002, \mn@doi [\mnras] {10.1046/j.1365-8711.2002.05013.x}, \href
  {https://ui.adsabs.harvard.edu/abs/2002MNRAS.329..597M} {329, 597 (MR02)}

\bibitem[\protect\citeauthoryear{{Nauenberg}}{{Nauenberg}}{1972}]{nauenberg-72}
{Nauenberg} M.,  1972, \mn@doi [\apj] {10.1086/151568}, \href
  {https://ui.adsabs.harvard.edu/abs/1972ApJ...175..417N} {175, 417}

\bibitem[\protect\citeauthoryear{{Norton}, {Wynn}  \& {Somerscales}}{{Norton}
  et~al.}{2004}]{norton-2004}
{Norton} A.~J.,  {Wynn} G.~A.,   {Somerscales} R.~V.,  2004, \mn@doi [\apj]
  {10.1086/423333}, \href
  {https://ui.adsabs.harvard.edu/abs/2004ApJ...614..349N} {614, 349}

\bibitem[\protect\citeauthoryear{{Patterson}}{{Patterson}}{1991}]{patterson-1991}
{Patterson} J.,  1991, \mn@doi [\pasp] {10.1086/132935}, \href
  {https://ui.adsabs.harvard.edu/abs/1991PASP..103.1149P} {103, 1149}

\bibitem[\protect\citeauthoryear{{Patterson}}{{Patterson}}{1994}]{patterson-1994}
{Patterson} J.,  1994, \mn@doi [\pasp] {10.1086/133375}, \href
  {https://ui.adsabs.harvard.edu/abs/1994PASP..106..209P} {106, 209}

\bibitem[\protect\citeauthoryear{{Pecaut} \& {Mamajek}}{{Pecaut} \&
  {Mamajek}}{2013}]{pecaut-13}
{Pecaut} M.~J.,  {Mamajek} E.~E.,  2013, \mn@doi [\apjs]
  {10.1088/0067-0049/208/1/9}, \href
  {https://ui.adsabs.harvard.edu/abs/2013ApJS..208....9P} {208, 9}

\bibitem[\protect\citeauthoryear{{Pottasch}}{{Pottasch}}{1959}]{pottasch-59}
{Pottasch} S.,  1959, Annales d'Astrophysique, \href
  {https://ui.adsabs.harvard.edu/abs/1959AnAp...22..412P} {22, 412}

\bibitem[\protect\citeauthoryear{{Reinsch}}{{Reinsch}}{1994}]{reinsch-94}
{Reinsch} K.,  1994, \aap, \href
  {https://ui.adsabs.harvard.edu/abs/1994A&A...281..108R} {281, 108}

\bibitem[\protect\citeauthoryear{{Revnivtsev}, {Churazov}, {Postnov}  \&
  {Tsygankov}}{{Revnivtsev} et~al.}{2009}]{revnivtsev-2009}
{Revnivtsev} M.,  {Churazov} E.,  {Postnov} K.,   {Tsygankov} S.,  2009,
  \mn@doi [\aap] {10.1051/0004-6361/200912317}, \href
  {https://ui.adsabs.harvard.edu/abs/2009A&A...507.1211R} {507, 1211}

\bibitem[\protect\citeauthoryear{{Revnivtsev}, {Potter}, {Kniazev}, {Burenin},
  {Buckley}  \& {Churazov}}{{Revnivtsev} et~al.}{2011}]{revnivtsev-2011}
{Revnivtsev} M.,  {Potter} S.,  {Kniazev} A.,  {Burenin} R.,  {Buckley}
  D.~A.~H.,   {Churazov} E.,  2011, \mn@doi [\mnras]
  {10.1111/j.1365-2966.2010.17765.x}, \href
  {https://ui.adsabs.harvard.edu/abs/2011MNRAS.411.1317R} {411, 1317}

\bibitem[\protect\citeauthoryear{{Ricker} et~al.,}{{Ricker}
  et~al.}{2015}]{ricker-2015}
{Ricker} G.~R.,  et~al., 2015, \mn@doi [Journal of Astronomical Telescopes,
  Instruments, and Systems] {10.1117/1.JATIS.1.1.014003}, \href
  {https://ui.adsabs.harvard.edu/abs/2015JATIS...1a4003R} {1, 014003}

\bibitem[\protect\citeauthoryear{{Sabbadin} \& {Bianchini}}{{Sabbadin} \&
  {Bianchini}}{1983}]{sabbadin-83}
{Sabbadin} F.,  {Bianchini} A.,  1983, \aaps, \href
  {https://ui.adsabs.harvard.edu/abs/1983A&AS...54..393S} {54, 393}

\bibitem[\protect\citeauthoryear{{Scott}, {Rawlings}  \& {Evans}}{{Scott}
  et~al.}{1994}]{scott-94}
{Scott} A.~D.,  {Rawlings} J.~M.~C.,   {Evans} A.,  1994, \mn@doi [\mnras]
  {10.1093/mnras/269.3.707}, \href
  {https://ui.adsabs.harvard.edu/abs/1994MNRAS.269..707S} {269, 707}

\bibitem[\protect\citeauthoryear{{Seaquist}, {Bode}, {Frail}, {Roberts},
  {Evans}  \& {Albinson}}{{Seaquist} et~al.}{1989}]{seaquist-89}
{Seaquist} E.~R.,  {Bode} M.~F.,  {Frail} D.~A.,  {Roberts} J.~A.,  {Evans} A.,
    {Albinson} J.~S.,  1989, \mn@doi [\apj] {10.1086/167846}, \href
  {https://ui.adsabs.harvard.edu/abs/1989ApJ...344..805S} {344, 805}

\bibitem[\protect\citeauthoryear{{Shahbaz}, {Watson}  \& {Dhillon}}{{Shahbaz}
  et~al.}{2014}]{shahbaz-14}
{Shahbaz} T.,  {Watson} C.~A.,   {Dhillon} V.~S.,  2014, \mn@doi [\mnras]
  {10.1093/mnras/stu267}, \href
  {https://ui.adsabs.harvard.edu/abs/2014MNRAS.440..504S} {440, 504}

\bibitem[\protect\citeauthoryear{{Shara}, {Prialnik}, {Hillman}  \&
  {Kovetz}}{{Shara} et~al.}{2018}]{shara-18}
{Shara} M.~M.,  {Prialnik} D.,  {Hillman} Y.,   {Kovetz} A.,  2018, \mn@doi
  [\apj] {10.3847/1538-4357/aabfbd}, \href
  {https://ui.adsabs.harvard.edu/abs/2018ApJ...860..110S} {860, 110}

\bibitem[\protect\citeauthoryear{{Steeghs} \& {Jonker}}{{Steeghs} \&
  {Jonker}}{2007}]{steeghs-2007}
{Steeghs} D.,  {Jonker} P.~G.,  2007, \mn@doi [\apjl] {10.1086/523848}, \href
  {https://ui.adsabs.harvard.edu/abs/2007ApJ...669L..85S} {669, L85}

\bibitem[\protect\citeauthoryear{{Stockman}, {Schmidt}, {Berriman}, {Liebert},
  {Moore}  \& {Wickramasinghe}}{{Stockman} et~al.}{1992}]{stockman-92}
{Stockman} H.~S.,  {Schmidt} G.~D.,  {Berriman} G.,  {Liebert} J.,  {Moore}
  R.~L.,   {Wickramasinghe} D.~T.,  1992, \mn@doi [\apj] {10.1086/172091},
  \href {https://ui.adsabs.harvard.edu/abs/1992ApJ...401..628S} {401, 628}

\bibitem[\protect\citeauthoryear{{Suleimanov}, {Revnivtsev}  \&
  {Ritter}}{{Suleimanov} et~al.}{2005}]{suleimanov-2005}
{Suleimanov} V.,  {Revnivtsev} M.,   {Ritter} H.,  2005, \mn@doi [\aap]
  {10.1051/0004-6361:20041283}, \href
  {https://ui.adsabs.harvard.edu/abs/2005A&A...435..191S} {435, 191}

\bibitem[\protect\citeauthoryear{{Suleimanov}, {Doroshenko}, {Ducci}, {Zhukov}
  \& {Werner}}{{Suleimanov} et~al.}{2016}]{suleimanov-16}
{Suleimanov} V.,  {Doroshenko} V.,  {Ducci} L.,  {Zhukov} G.~V.,   {Werner} K.,
   2016, \mn@doi [\aap] {10.1051/0004-6361/201628301}, \href
  {https://ui.adsabs.harvard.edu/abs/2016A&A...591A..35S} {591, A35}

\bibitem[\protect\citeauthoryear{{Suleimanov}, {Doroshenko}  \&
  {Werner}}{{Suleimanov} et~al.}{2019}]{suleimanov-19}
{Suleimanov} V.~F.,  {Doroshenko} V.,   {Werner} K.,  2019, \mn@doi [\mnras]
  {10.1093/mnras/sty2952}, \href
  {https://ui.adsabs.harvard.edu/abs/2019MNRAS.482.3622S} {482, 3622}

\bibitem[\protect\citeauthoryear{{Torres}, {Casares}, {Jim{\'e}nez-Ibarra},
  {{\'A}lvarez-Hern{\'a}ndez}, {Mu{\~n}oz-Darias}, {Armas Padilla}, {Jonker}
  \& {Heida}}{{Torres} et~al.}{2020}]{torres-2020}
{Torres} M.~A.~P.,  {Casares} J.,  {Jim{\'e}nez-Ibarra} F.,
  {{\'A}lvarez-Hern{\'a}ndez} A.,  {Mu{\~n}oz-Darias} T.,  {Armas Padilla} M.,
  {Jonker} P.~G.,   {Heida} M.,  2020, \mn@doi [\apjl]
  {10.3847/2041-8213/ab863a}, \href
  {https://ui.adsabs.harvard.edu/abs/2020ApJ...893L..37T} {893, L37}

\bibitem[\protect\citeauthoryear{{Wada}, {Yuasa}, {Nakazawa}, {Makishima},
  {Hayashi}  \& {Ishida}}{{Wada} et~al.}{2018}]{wada-2018}
{Wada} Y.,  {Yuasa} T.,  {Nakazawa} K.,  {Makishima} K.,  {Hayashi} T.,
  {Ishida} M.,  2018, \mn@doi [\mnras] {10.1093/mnras/stx2880}, \href
  {https://ui.adsabs.harvard.edu/abs/2018MNRAS.474.1564W} {474, 1564}

\bibitem[\protect\citeauthoryear{{Warner}}{{Warner}}{1995}]{warner-libro}
{Warner} B.,  1995, Cambridge Astrophysics Series, \href
  {https://ui.adsabs.harvard.edu/abs/1995CAS....28.....W} {28}

\bibitem[\protect\citeauthoryear{{Watson}, {King}  \& {Osborne}}{{Watson}
  et~al.}{1985}]{watson-1985}
{Watson} M.~G.,  {King} A.~R.,   {Osborne} J.,  1985, \mn@doi [\mnras]
  {10.1093/mnras/212.4.917}, \href
  {https://ui.adsabs.harvard.edu/abs/1985MNRAS.212..917W} {212, 917}

\bibitem[\protect\citeauthoryear{{Webbink}, {Rappaport}  \&
  {Savonije}}{{Webbink} et~al.}{1983}]{webbink-83}
{Webbink} R.~F.,  {Rappaport} S.,   {Savonije} G.~J.,  1983, \mn@doi [\apj]
  {10.1086/161159}, \href
  {https://ui.adsabs.harvard.edu/abs/1983ApJ...270..678W} {270, 678}

\bibitem[\protect\citeauthoryear{{Webbink}, {Livio}, {Truran}  \&
  {Orio}}{{Webbink} et~al.}{1987}]{webbink-87}
{Webbink} R.~F.,  {Livio} M.,  {Truran} J.~W.,   {Orio} M.,  1987, \mn@doi
  [\apj] {10.1086/165095}, \href
  {https://ui.adsabs.harvard.edu/abs/1987ApJ...314..653W} {314, 653}

\bibitem[\protect\citeauthoryear{{Williams}}{{Williams}}{1901}]{williams-1901}
{Williams} A.~S.,  1901, \mn@doi [\mnras] {10.1093/mnras/61.5.337}, \href
  {https://ui.adsabs.harvard.edu/abs/1901MNRAS..61..337W} {61, 337}

\bibitem[\protect\citeauthoryear{{Yaron}, {Prialnik}, {Shara}  \&
  {Kovetz}}{{Yaron} et~al.}{2005}]{yaron-2005}
{Yaron} O.,  {Prialnik} D.,  {Shara} M.~M.,   {Kovetz} A.,  2005, \mn@doi
  [\apj] {10.1086/428435}, \href
  {https://ui.adsabs.harvard.edu/abs/2005ApJ...623..398Y} {623, 398}

\bibitem[\protect\citeauthoryear{{Yee}, {Petigura}  \& {von Braun}}{{Yee}
  et~al.}{2017}]{yee-2017}
{Yee} S.~W.,  {Petigura} E.~A.,   {von Braun} K.,  2017, \mn@doi [\apj]
  {10.3847/1538-4357/836/1/77}, \href
  {https://ui.adsabs.harvard.edu/abs/2017ApJ...836...77Y} {836, 77}

\bibitem[\protect\citeauthoryear{{Zi{\'o}{\l}kowski} \&
  {Zdziarski}}{{Zi{\'o}{\l}kowski} \& {Zdziarski}}{2020}]{ziolkowski-2020}
{Zi{\'o}{\l}kowski} J.,  {Zdziarski} A.~A.,  2020, \mn@doi [\mnras]
  {10.1093/mnras/staa3088}, \href
  {https://ui.adsabs.harvard.edu/abs/2020MNRAS.499.4832Z} {499, 4832}

\bibitem[\protect\citeauthoryear{{Zorotovic}, {Schreiber}  \&
  {G{\"a}nsicke}}{{Zorotovic} et~al.}{2011}]{zorotovic-2011}
{Zorotovic} M.,  {Schreiber} M.~R.,   {G{\"a}nsicke} B.~T.,  2011, \mn@doi
  [\aap] {10.1051/0004-6361/201116626}, \href
  {https://ui.adsabs.harvard.edu/abs/2011A&A...536A..42Z} {536, A42}

\bibitem[\protect\citeauthoryear{{{\v{S}}imon}}{{{\v{S}}imon}}{2002}]{simon-2002}
{{\v{S}}imon} V.,  2002, \mn@doi [\aap] {10.1051/0004-6361:20011560}, \href
  {https://ui.adsabs.harvard.edu/abs/2002A&A...382..910S} {382, 910}

\makeatother
\end{thebibliography}


\bsp	
\label{lastpage}
\end{document}